\documentclass[oneside]{article}
\usepackage[T1]{fontenc}
\usepackage{authblk}
\usepackage{float}
\usepackage{amsmath}
\usepackage{graphicx}
\usepackage{xfrac}
\usepackage[english]{babel}
\usepackage{lmodern}
\usepackage[T1]{fontenc}
\usepackage[latin9]{inputenc}

\usepackage{csquotes}

\usepackage{longtable}

\usepackage{mathtools}
\usepackage{graphicx}
\usepackage{esint}
\usepackage[unicode=true,
 bookmarks=false,
breaklinks=false,pdfborder={0 0 1},backref=false,colorlinks=false]
 {hyperref}
\usepackage[a4paper, total={210mm,297mm},margin=2.0cm]{geometry}

\providecommand{\tabularnewline}{\\}

\title{LBsoft: a parallel open-source software for simulation of colloidal systems}

\author[1,2]{F. Bonaccorso}

\author[1]{A. Montessori}

\author[1,2]{A. Tiribocchi}

\author[4]{G. Amati}

\author[1]{M. Bernaschi}

\author[1]{M. Lauricella\thanks{Electronic address: \texttt{m.lauricella@iac.cnr.it}; Corresponding author}}

\author[1,2,3]{S. Succi}

\affil[1]{Istituto per le Applicazioni del Calcolo CNR, via dei Taurini 19, Rome, Italy}
\affil[2]{Center for Life Nano Science@La Sapienza, Istituto Italiano di Tecnologia, 00161 Roma, Italy}
\affil[3]{Institute for Applied Computational Science, John A. Paulson School of Engineering and Applied Sciences, Harvard University, Cambridge, USA}
\affil[4]{SCAI, SuperComputing Applications and Innovation Department, CINECA, Via dei Tizii, 6, Rome 00185, Italy}

\date{\today}

\begin{document}

\maketitle

\begin{abstract}
We present LBsoft, an open-source software developed mainly to simulate the hydro-dynamics 
of colloidal systems based on the concurrent coupling between lattice Boltzmann methods for the fluid and 
discrete particle dynamics for the colloids.
Such coupling has been developed before, but, to the best of our knowledge, no detailed discussion of the
programming issues to be faced in order to attain efficient implementation on parallel architectures, has ever
been presented to date.   
In this paper, we describe in detail the underlying multi-scale models, their coupling procedure, along side with a 
description of the relevant input variables, to facilitate third-parties usage.
  
The code is designed to exploit parallel computing platforms, taking advantage also of the 
recent AVX-512 instruction set.  We focus on LBsoft structure, functionality, parallel 
implementation, performance and availability, so as to facilitate the access to 
this computational tool to the research community in the field.   

The capabilities of LBsoft are highlighted for a number of prototypical case studies, such 
as pickering emulsions, bicontinuous systems, as well as an original study 
of the coarsening process in confined bijels under shear.
 

\vspace{0.2cm}

\textbf{PROGRAM SUMMARY}
\vspace{0.2cm}

\begin{small}
\noindent
{\em Program Title:} LBsoft                                     \\
{\em Licensing provisions:} 3-Clause BSD License                                   \\
{\em Programming language:} Fortran 95                                  \\
{\em Nature of problem:} Hydro-dynamics of the colloidal multi-component systems and Pickering emulsions.\\
{\em Solution method:} Numerical solutions to the Navier-Stokes equations by Lattice-Boltzmann (lattice-Bhatnagar-Gross-Krook, LBGK) method [1] describing the fluid dynamics within an Eulerian description. Numerical solutions to the equations of motion describing a set of discrete colloidal particles within a Lagrangian representation coupled to the LBGK solver [2]. The numerical solution of the coupling algorithm includes the back reaction effects for each force terms following a multi-scale paradigm.\\
   \\

[1] S. Succi, The Lattice Boltzmann Equation: For Fluid Dynamics and Beyond, Oxford University Press, 2001.         

[2] A. Ladd,R. Verberg, Lattice-Boltzmann simulations of particle-fluid suspensions, J. Stat. Phys. 104.5-6 (2001) 1191-1251
\end{small}

\end{abstract}

\section{Introduction}
\label{Introd}

In the last two decades, the design of novel mesoscale soft materials has gained considerable interest. 
In particular, soft-glassy materials (SGM) like emulsions, foams and gels have received a growing attention 
due to their applications in several areas of modern industry.
For instance, chemical and food processing, manufacturing and biomedical involve several soft-glassy materials nowadays \cite{fernandez2016fluids,piazza2011soft,mezzenga2005understanding}.
Besides their technological importance, the intriguing non-equilibrium effects, 
such as long-time relaxation, yield-stress behaviour and highly non-Newtonian 
dynamics make these materials of significant theoretical interest. 
A reliable model of these features alongside with its software implementation is of compelling interest for the rational designing and shaping up of novel soft porous materials. In this context, computational fluid dynamics (CFD)  
can play a considerable role in numerical investigations of the underpinning physics 
in order to enhance mainly mechanical properties, thereby opening new scenarios 
in the realisation of new states of matter.

Nowadays, the lattice-Boltzmann method (LBM) is one of the most popular techniques for the simulation of fluid dynamics due to its relative simplicity and locality of the underlying algorithm.  
A straightforward parallelization of the method is one of the main advantages of the lattice-Boltzmann algorithm which makes it an excellent tool for high-performance CFD. 

Starting with the LBM 
development, a considerable effort has been carried out to include the hydrodynamic interactions 
among solid particles and fluids \cite{ladd2015lattice,ladd2001lattice,aidun1998direct,
ladd1994numericala}. This extension has opened the possibility to simulate complex colloidal systems also referred to as Pickering emulsions \cite{pickering1907cxcvi}. In this framework, the description of the colloidal particles exploits a Lagrangian dynamics solver coupled to the lattice-Boltzmann solver of fluid dynamics equations. This intrinsically multiscale approach 
can catch the dynamical transition from a bi-continuous interfacially jammed emulsion gel to 
a Pickering emulsion, highlighting the mechanical and spatial properties which are of primary 
interest for the rational design of SGM \cite{xie2017direct,liu2016multiphase,
frijters2012effects,jansen2011bijels}.

By now, several LBM implementations are available, both academic packages 
such as OpenLB \cite{heuveline2007openlb}, DL\_MESO \cite{seaton2013dl_meso}, 
also implementing particle-fluid interaction such 
as WaLBerla \cite{bauer2020walberla,schornbaum2016massively}, Palabos \cite{latt2009palabos}, Ludwig \cite{desplat2001ludwig}, HemeLB \cite{mazzeo2008hemelb}, and commercially 
licensed software such as XFlow \cite{holman2012solution} and PowerFLOW \cite{lockard2002evaluation}, 
to name a few. 
The modelling of SGM requires specific implementations of LBM and Lagrangian solvers aimed at the optimization of the multi-scale coupling algorithms in order to scale up to large systems. 
This point is mandatory for an accurate statistical description of such complex systems 
which is necessary for a fully rational design of a new class of porous materials.

In this paper, we present, along with the overall model, LBsoft, a parallel FORTRAN code, specifically designed to simulate bi-continuous systems with colloidal particles under a variety of different conditions. This comprehensive platform is devised in such a way to handle both LBGK solver and particle dynamics via a scalable multiscale algorithm. The framework is developed to exploit several computational architectures and parallel decompositions. With most of the parameters taken from relevant literature in the field, several test cases have been carried out as a software benchmark.

The paper is structured as follows. In Section 2 we report a brief description of the underlying method.  A full derivation of the model is not reported, but we limit the presentation to the main ingredients necessary to describe the implementation of the algorithms. Nonetheless,  original references of models are reported step by step along with the implementation description. In Section 3 we describe the details on memory organisation and parallel communication patterns. In Section 4 we report a set of tests used to validate the implementation and assess the parallel performance. Finally, conclusions and an outlook on future development directions are presented.

\section{Method}

In this section, we briefly review the methodology implemented in LBsoft. Further details are available in the literature.

The code combines two different levels of description: the first exploits 
a continuum description for the hydrodynamics of a single or two immiscible fluids; the second utilises individual 
rigid bodies for the representation of colloidal particles or other suspended species. 
The exchange of information between the two levels is computed on-the-fly, allowing a simultaneous 
evolution in time of both fluids and particles.

\subsection{Single component lattice-Boltzmann}

\begin{figure}
\label{figd3q19}
\begin{center}
\centerline{\resizebox{.5\linewidth}{!}{\includegraphics{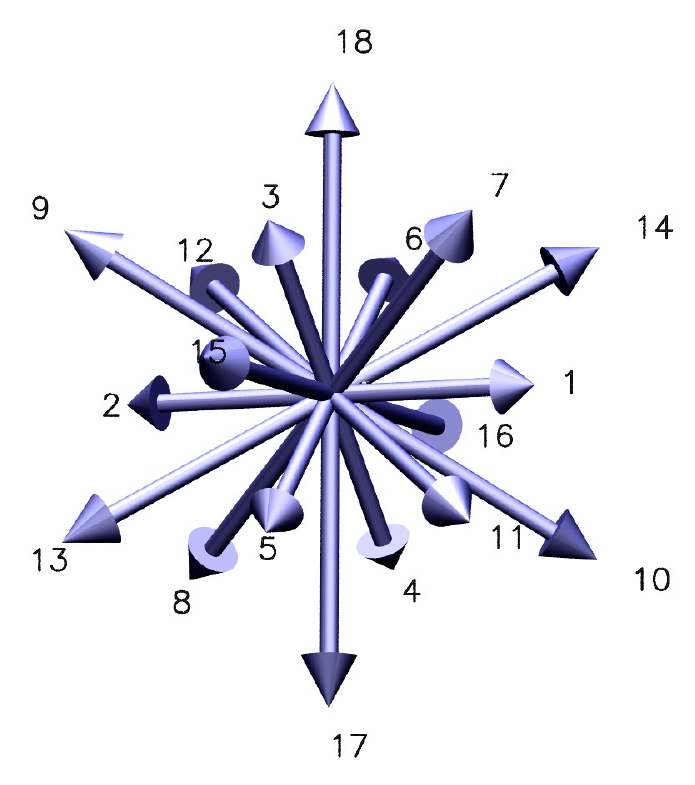}}}
\end{center}
\caption{The set of discrete velocities for the three-dimensional 19-speed cubic 
lattice (D3Q19).
}
\end{figure}

The lattice-Boltzmann equation facilitates the simulation of flows and 
hydrodynamic interactions in fluids by a fully discretised analogue of the Boltzmann kinetic equation.
In the LBM approach, the fundamental quantity is $f_{i}(\vec{r};t)$, namely the probability of finding a 
``fluid particle'' at the spatial mesh point $\vec{r}$ and at time $t$ with velocity $c_i$ selected from a 
finite set of possible speeds. The LBsoft code implements the three-dimensional 19-speed cubic 
lattice (D3Q19) with the discrete velocities $c_i$ ($i \in [0,...,18]$) connecting mesh points with spacing $\Delta x$ , 
located at distance $1\Delta x$ and $\sqrt 2 \Delta x$, first and second neighbors, respectively (see Fig. \ref{figd3q19}).

Denoted $\rho(\vec{r};t)$ and $\vec{u}(\vec{r};t)$ respectively 
the local density and velocity, 
the lattice-Boltzmann equation is implemented in 
single-relaxation time 
(Bhatnagar-Gross-Krook equation) as follows
\begin{equation}
\label{eq:bgk}
f_{i}(\vec{r}+\vec{c}_{i};t+1)=(1-\omega)f_{i}(\vec{r};t)+\omega f_{i}^{eq}(\rho(\vec{r};t),\vec{u}(\vec{r};t))
\end{equation}
where $f^{eq}$ is the lattice local equilibrium, basically the 
local Maxwell-Boltzmann 
distribution, and $\omega$ is a frequency tuning the relaxation 
towards the local equilibrium on a
timescale $\tau=1/\omega$. Hereinafter, the right-hand side of 
Eq. \ref{eq:bgk} is refereed to as the post-collision population 
and indicated by the symbol $f^*_{i}(\vec{r};t)$, so that 
Eq. \ref{eq:bgk} can be also written in the following two-step 
form: the collision step given by
\begin{equation}
\label{eq:bgk1}
f^*_{i}(\vec{r};t)=(1-\omega)f_{i}(\vec{r};t)+\omega f_{i}^{eq}(\rho(\vec{r};t),\vec{u}(\vec{r};t)),
\end{equation}
and a subsequent streaming step:
\begin{equation}
\label{eq:bgk2}
f_{i}(\vec{r}+\vec{c}_{i};t+1)=f^*_{i}(\vec{r};t).
\end{equation}
Note that the relaxation frequency $\omega$ controls the kinematic viscosity of the fluids by the relation
\begin{equation}
\nu=c_{s}^{2}\frac{\Delta x^{2}}{\Delta t} \left(\frac{1}{\omega}-\frac{1}{2}\right)\label{eq:viscosity}
\end{equation}
where $\Delta x$ and $\Delta t$ are the physical length and 
time of the correspondent counterparts in lattice units. Moreover, the positivity of the kinematic viscosity
imposes the condition $0<\omega<2$.
In the actual implementation, the equilibrium function is computed as an expansion of the Maxwell-Boltzmann 
distribution in the Mach number
truncated to the second order
\begin{equation}
f_{i}^{eq}=w_{i}\rho(1+u_{i}+\frac{1}{2}q_{i})\label{eq:LEQ}
\end{equation}
where $u_{i}=c_{ia}u_{a}/c_{s}^{2}$ and $q_{i}=(c_{ia}c_{ib}-c_{s}^{2}\delta_{ab})u_{a}u_{b}/c_{s}^{4}$
denote the dipole and quadrupole flow contributions, respectively, whereas $w_{i}$ is a set
of lattice weights which are
the lattice counterpart of the global Maxwell-Boltzmann distribution in absence of flows,
in the actual lattice: $w_0=1/3$, $w_{1-6}=1/18$, and $w_{7-18}=1/36$ \cite{kruger2017lattice}. 
Note that $c_{s}^{2}=\sum_{i}w_{i}c_{ia}^{2}$ is the square lattice sound speed equal to 1/3 in lattice units.
Given the above prescriptions, the local hydrodynamic variables are computed as:
\begin{equation} \label{eq:rho}
\rho = \sum_i f_{i}\left(\vec{r},\,t\right) 
\end{equation}
and 
\begin{equation} \label{eq:vel}
\rho \vec{u} = \sum_i f_{i}\left(\vec{r},\,t\right) \vec{c}_{i}
\end{equation}

with the lattice fluid
obeying an ideal equation of state $p=\rho c_{s}^{2}$.

\subsection{Boundary conditions}

LBsoft supports a number of different conditions for the populations 
of fluid nodes close to the boundaries of the mesh. In particular, LBsoft exploits the simple bounce-back 
rule and its modified versions to treat Dirichlet and Neumann boundary conditions, 
enabling fluid inlet and outlet open boundaries.
The bounce-back rule allows to achieve zero velocity 
along a link, $\vec{c}_i$, hitting an obstacle from a fluid node located at $\vec{r}_b$, in the following called boundary node.
Briefly, the velocity, $\vec{c}_i$, of the incoming $i$-th populations is reverted at 
the wall to the opposite direction, $\vec{c}_{\bar{i}}=-\vec{c}_i$. The wall location is assumed 
at halfway, $\vec{r}_w=\vec{r}_b+\sfrac{1}{2}\vec{c}_i$, between the boundary node,
$\vec{r}_b$ and solid node, $\vec{s}=\vec{r}_b+\vec{c}_i$, 
so the method is also referred to as the halfway bounce-back.
The zero velocity bounce-back rule \cite{succi2018lattice,kruger2017lattice} is realized through a variant of the 
streaming step in Eq. \ref{eq:bgk2}:
\begin{equation} 
\label{eq:bb}
f_{\bar{i}}(\vec{r}_b;t+1)=f^*_{i}(\vec{r}_b;t).
\end{equation}

A simple correction of Eq. \ref{eq:bb}
provides the extension to the Neumann boundary condition (velocity constant) \cite{ladd2001lattice,ladd1994numericalb}:
\begin{equation} 
\label{eq:bbv}
f_{\bar{i}}(\vec{r}_b;t+1)=f^*_{i}(\vec{r}_b;t) -2 w_i \rho_w \frac{\vec{u}_w \cdot \vec{c}_i}{c^2_s},
\end{equation}
where $\vec{u}_w$ denotes the fluid velocity and $\rho_w$ the density at the inlet
location, $\vec{r}_w=\vec{r}+\sfrac{1}{2}\vec{c}_i$, estimated by extrapolation. 
For example, given $\vec{r}_{b+1}$, the locations of the next interior node, $\rho_w$ 
can be assessed as $\rho_w=\rho_w(\vec{r}_b)+\sfrac{1}{2}[\rho_w(\vec{r}_b)-\rho_w(\vec{r}_{b+1})]$ or
by means of a finite difference scheme with greater accuracy.

A slightly different bounce-back method is used to reinforce the prescription of 
pressure (Dirichlet condition) at open boundaries, which is usually referred to 
as the anti-bounce-back method \cite{kruger2017lattice,ginzburg2008two}.
In particular, the sign of the reflected population changes. Hence, the distribution 
is modified as
\begin{equation} 
\label{eq:bbp}
f_{\bar{i}}(\vec{r}_b;t+1)=-f^*_{i}(\vec{r}_b;t) + 2 w_i \rho_w \left[1+\frac{(\vec{u}_w \cdot \vec{c}_i)^2}{2c^4_s}-\frac{\vec{u}_w^2}{2c^2_s}\right],
\end{equation}
$\rho_w$ being an external parameter. 
As in the previous case, the value of $\vec{u}_w$ can be extrapolated by following a finite difference based approach,
$\vec{u}_w=\vec{u}_w(\vec{r}_b)+\sfrac{1}{2}[\vec{u}_w(\vec{r}_b)-\vec{u}_w(\vec{r}_{b+1})]$.

\subsection{Two--component lattice-Boltzmann \label{twocomp}}
The Shan-Chen approach provides a straightforward way
to include a non-ideal term in the equation of state of a two-component system.
If the two components are denoted $k$ and $\bar{k}$, 
the basic idea is to add an extra cohesive force usually defined as
\begin{equation}\label{shanchen}
\vec{F}^{k}(\vec{r},t)=\psi^k(\vec{r},t)G_C\sum_{i}w_{i}\psi^{\bar{k}}(\vec{r}+\vec{c}_{i},t)\vec{c_{i}},
\end{equation}
acting on each component.
In Eq \ref{shanchen}, $G_C$ is a parameter tuning the strength of the inter-component force, and
$\psi(\rho)$ is a function of local density playing the role of an effective density.
For sake of simplicity, we adopt $\psi^k(\vec{r},t)=\rho^k(\vec{r},t)$, given $\rho^k(\vec{r},t)$ 
the local density of the k$-th$ component. Therefore, $\psi$ will be replaced by $\rho$ in the following text.
This extra force term delivers a non-ideal equation of state having the following
form 
\begin{equation}
p=c_{s}^{2}(\rho^{k}+\rho^{\bar{k}})+c_{s}^{2} G_C \rho^{k} \rho^{\bar{k}}. \label{SCEOS}
\end{equation}
Note that the fluid-fluid surface tension $\sigma$ is subordinated on the parameter $G_C$.
Further, it is possible to estimate a critical value $G_{C,crit}=\frac{1}{\rho^{k}+\rho^{\bar{k}}}$
at which the system exhibits a mixture demixing in two distinct fluid domains,
becoming increasingly ``pure'' as the parameter $G_C$ increases.
The extra force term is included in the LBM by
a source term $S^k$ added to Eq \ref{eq:bgk} for each k$-th$ component 
\begin{equation}\label{eq:twobgk}
f^k_{i}(\vec{r}+\vec{c}_{i};t+1)=(1-\omega_{eff})f^k_{i}(\vec{r};t)+\omega_{eff} f_{i}^{eq}(\rho^k(\vec{r};t),\vec{u}(\vec{r};t))+S^k_{i}(\vec{r};t)
\end{equation}
where $\omega_{eff}=2c_s^2/(2 \bar{\nu} -c_s^2)$ is related to the kinematic 
viscosity $\bar{\nu}$ of the mixture computed as
$\frac{1}{\bar{\nu}}=\frac{\rho_k}{(\rho_k+\rho_{\bar{k}})}\frac{1}{\nu_k} + \frac{\rho_{\bar{k}}}{(\rho_k+\rho_{\bar{k}})}\frac{1}{\nu_{\bar{k}}}$ where $\nu_k$ and $\nu_{\bar{k}}$ denote the kinematic viscosities of the two fluids, taken individually.
Given $\rho^k = \sum_i f^k_{i}$ and $\rho^k \vec{u}^k = \sum_i f^k_{i} \vec{c}_{i}$, 
the total fluid density is simply $\rho = \sum_k  \rho^{k}$,
whereas the effective velocity of the mixture is
computed as 
\begin{equation}\label{eq:scvel}
\vec{u} = \frac{\sum_k \omega^k  \rho^{k} \vec{u}^k}{\sum_k  \omega^k \rho^{k}}.
\end{equation}
The extra term $S^k_{i}(\vec{r};t)$ is computed by the Exact Difference Method, proposed by Kupershtokh {\em et al.}, as the difference between the lattice local equilibrium at a shifted fluid velocity and the equilibrium at effective velocity of the mixture
\begin{equation}
S^k_{i}(\vec{r};t)=f_{i}^{eq}(\rho^k,\vec{u}+\frac{\vec{F}^{k} \Delta t}{\rho^k})-f_{i}^{eq}(\rho^k,\vec{u}).\label{eq:edm}
\end{equation}
The equilibrium distribution functions $f_i^{eq}$ for the k$-th$ component are still computed by Eq \ref{eq:LEQ} with density $\rho^k$ and the velocity $\vec{u}$ obtained from Eq \ref{eq:scvel}.
By inserting Eq \ref{eq:edm} in \ref{eq:twobgk} we obtain the complete BGK equation 
\begin{equation}\label{eq:lbsbgk}
\begin{split}
f^k_{i}(\vec{r}+\vec{c}_{i};t+1)=(1-\omega_{eff})[f^k_{i}(\vec{r};t)-f_{i}^{eq}(\rho^k(\vec{r};t),\vec{u}(\vec{r};t))]+ \\
+f_{i}^{eq}(\rho^k,\vec{u}+\frac{\vec{F}^{k} \Delta t}{\rho^k})
\end{split}
\end{equation}
in the form actually implemented in LBsoft. Note that, in this forcing scheme, the correct momentum flux is computed at half time step as $\rho^k \vec{u}^k = \sum_i f^k_{i} \vec{c}_{i}+\sfrac{1}{2}\vec{F}^{k} \Delta t$, which differs from the velocity inserted in the lattice equilibrium function.

\subsection{Colloidal rigid particles \label{refmd}}
\begin{figure}
\label{FigSphere}
\begin{center}
\centerline{\resizebox{.5\linewidth}{!}{\includegraphics{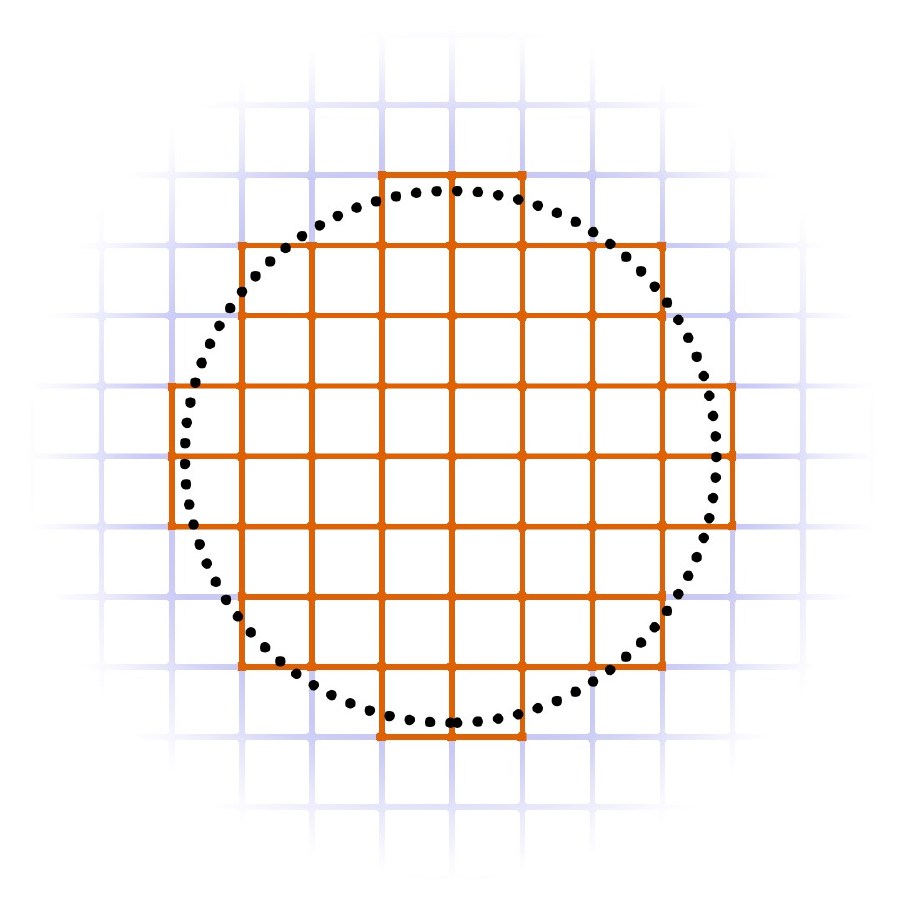}}}
\end{center}
\caption{Two dimensional sketch of a staircase approximation (orange solid line) of a spherical particle (dotted).
}
\end{figure}

The rigid body description involves a Lagrangian solver for the particle evolution following the trailblazing work by Tony Ladd \cite{ladd1994numericala, ladd1994numericalb}.
In the original Ladd's method, the solid particle (colloid) is represented by a closed surface ${\mathcal S}$, taken, for simplicity, as a rigid sphere in the following. To reduce the computational burden, we use a staircase approximation $\Sigma$ of the sphere ${\mathcal S}$, given as the set of lattice links cut by ${\mathcal S}$.
As a result, each particle $p$ is represented by a staircase sphere of surface $\Sigma_p$ (see Fig. \ref{FigSphere}).

The accuracy of this representation is only $O(1/R)$, $R$ being the 
radius of the sphere in lattice units, so that, in actual practice,
the simulations always deal with rough spheres.

In accordance with the formulation proposed by F. Jansen and J. Harting \cite{jansen2011bijels}, only the exterior 
regions are filled with fluid, whereas the interior parts of the particles are considered solid nodes. In the following, we designate as a boundary node, $\vec{r}_b$, all the fluid nodes linked by a vector $\vec{c}_i$ to a solid node, $\vec{r}_s=\vec{r}_b+\vec{c}_i$, of the particle.
The solid--fluid interaction proceeds as follows.
Since the particle surface (wall) of the staircase approximation is placed in the 
middle point of a lattice link, we identify pairs of mirror
directions, denoted $(i,\bar i)$, with $i$ oriented along the cut 
link hitting the particle surface and 
$\bar{i}$ directed in the opposite way, $\vec{c}_{\bar{i}}=-\vec{c}_{i}$.
Hence, the populations hitting the particle surface are managed via a simple
generalization of the halfway bounce-back rule 
\cite{kruger2017lattice} including the correction shown in Eq. \ref{eq:bbv} modelling
the relative motion of the solid particle with respect 
to the surrounding fluid medium.
Assuming a time step equal to one for simplicity,
a post-collision population, $f^{\star}_i(\vec{r}_b,t)$, hitting the wall along the cut 
link $i$ is reflected at the wall location,  $\vec{r}_{w} = \vec{r}_b + \frac {1}{2} \vec{c}_i$,   
along the cut link $\bar{i}$ following the rule:
\begin{equation}\label{eq:LADDFLBE}
f_{\bar{i}} \left(\vec{r}_b, t+1 \right) =f_{i}^{\star}  \left( \vec{r}_b, t \right) - 2 w_i \rho  \frac{\vec{u}_w \cdot \vec{c}_i}{c_s^2}.
\end{equation}
The symbol $\vec{u}_w$ denotes the wall velocity of the $p$-th particle given by
\begin{equation}
\label{UPI}
\vec{u}_{w} = \vec{v}_p + \left(\vec{r}_w-\vec{r}_p\right) \times \vec{\omega}_p.
\end{equation}
All coordinates with $p$-th as a subscript are relative to the center of the $p$-th particle, located at position $\vec{r}_p$ and moving with translation and angular velocities $\vec{v}_p$ and $\vec{\omega}_p$, respectively. Note that these rules reduce to the usual bounce-back conditions for a solid at rest, $\vec{v}_p=\vec{\omega}_p=0$.
As a result of the reflection rule in Eq. \ref{eq:LADDFLBE}, the force acting on the $p$-th particle
at the wall location, $\vec{r}_{w}$, is

\begin{equation}\label{eq:LADDFORCE}
\begin{split}
\vec{F}_{i} \left(\vec{r}_w, t+\frac{1}{2} \right) = 
\left[ f_{\bar{i}} \left(\vec{r}_b, t+1 \right) + f_{i}^{\star} \left( \vec{r}_b, t \right)  \right] \vec{c}_i  = \\
= \left[ 2 f_{i}^{\star}  \left( \vec{r}_b, t \right) - 2 w_i \rho  \frac{\vec{u}_w \cdot \vec{c}_i}{c_s^2} \right] \vec{c}_i .
\end{split}
\end{equation}

\begin{figure}
\label{FigLadd}
\begin{center}
\centerline{\resizebox{.5\linewidth}{!}{\includegraphics{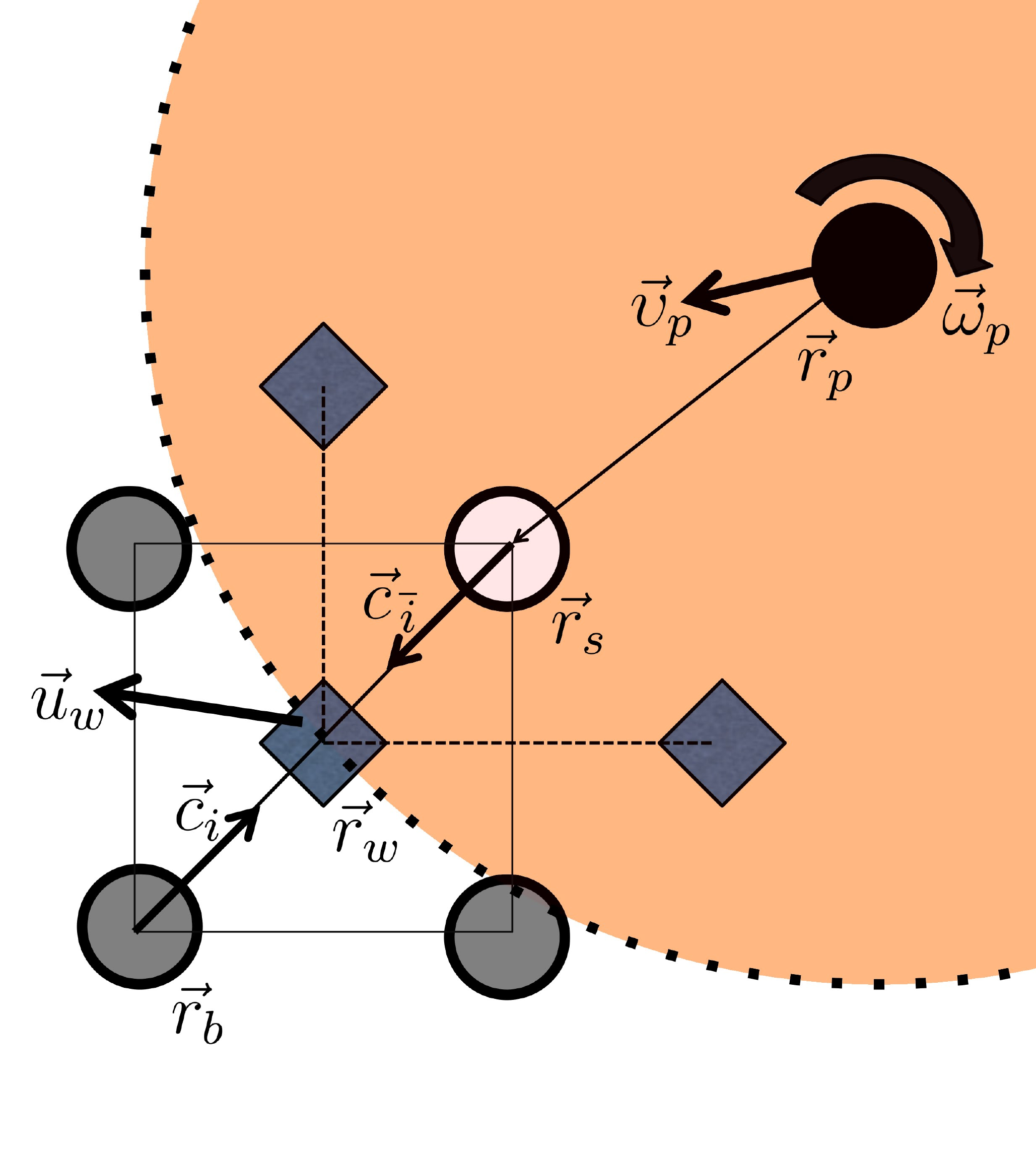}}}
\end{center}
\caption{Head-on collisions along a link cutting through the staircased surface (diamonds). The actual surface is represented by the dotted line.}
\end{figure}

The different wettability of the particle surface provides a global 
unbalance of forces and a corresponding torque on 
the rigid body which is accounted by a variant approach 
of the model for hydrophobic fluid-surface interactions originally introduced 
by Benzi et al. \cite{sbragaglia2006surface} and extended to the realm of particle 
simulations in following works \cite{jansen2011bijels,onishi2008lattice}.
In LBsoft, each solid node of 
the particle frontier is filled with a virtual fluid density. Hence, the virtual 
density at the node is taken as the average density value of the only fluid 
nodes $i_{\text{NP}}$ in the nearest lattice positions. 
In symbols, the virtual density of the $k$-th component reads 
\begin{equation}
\label{soliddens}
\bar{\rho}_s^k(\vec{r},t)=\left( 1+ \zeta_{ads}(\vec{r}) \right) \frac{\sum_{i_{\text{NP}}}w_{i}\rho^{k}(\vec{r}+\vec{c}_{i},t)}{\sum_{i_{\text{NP}}}w_{i}},
\end{equation}
where the average is weighted on the lattice stencil. Note that the average is multiplied by a prefactor
including the constant $\zeta_{ads}(\vec{r})$, which tunes the local wettability of the solid node placed in $\vec{r}$. For $\zeta_{ads}>0$ the particle surface prefers the $k$-th fluid component, as for $\zeta_{ads}<0$ \textit{vice versa}.
Hence, the Shan-Chen force acting on the solid node is
\begin{equation}\label{shanchen2}
\vec{F}^{k}_{s}(\vec{r},t)=\bar{\rho}^{k}_{s}(\vec{r},t)G_C\sum_{i}w_{i}\rho^{\bar{k}}(\vec{r}+\vec{c}_{i},t) s(\vec{r}+\vec{c}_{i},t) \vec{c_{i}},
\end{equation}
where $s$ denotes a switching function that is equal to 1 or 0 for a fluid or solid node, respectively.
The latter force term is counterbalanced by the Shan-Chen force exerted from the particle 
surface on the surrounding fluid, which is computed in Eq. \ref{shanchen} with the nodes filled with virtual fluid handled as regular fluid nodes.

The net force, $\vec{F}_{liq,\,p}$, acting from the surrounding fluid on the $p$-th particle is obtained by summing over all wall sites, $\vec{r}_w$, associated to the reflected links $i$, namely:
\begin{equation}
\label{FORCEPARTEQ}
\vec{F}_{liq,\,p} = \sum_{{\vec{r}_w,i} \in \Sigma_p} \vec{F}_{i}(\vec{r}_w).
\end{equation}
Similarly, the corresponding torque, $\vec{T}_{liq,\,p}$, is computed as
\begin{equation}
\label{TORQUEPARTEQ}
\vec{T}_{liq,\,p} = \sum_{{\vec{r}_w,i} \in \Sigma_p}  \left(\vec{r}_w-\vec{r}_p\right) \times \vec{F}_{i}(\vec{r}_w).
\end{equation}

Since particles move over the lattice nodes, it happens that a sub set of fluid boundary nodes, denoted $\{\vec{r}_d\}\in\mathcal{B}_p$, 
in front of a moving $p$-th particle cross its surface becoming solid nodes. Similarly, 
a sub set of interior nodes on the surface, $\{\vec{r}_c\}\in\Sigma_p$, are released
at the back of the $p$-th moving particle. The two distinct events require the destruction 
and the creation of fluid nodes, respectively. 
In the destruction fluid node step, LBsoft pursues the straightforward procedure
originally proposed by Aidun {\em et al.} \cite{aidun1998direct}. Whenever a fluid 
node changes to solid, the fluid
is deleted, and its linear and angular momenta are transferred to the particle. 
Given $\vec{r}_d$ the position vector of a deleted fluid node, the total force, $\vec{F}_{d,p}$,
and torque, $\vec{T}_{d,p}$, on the $p$-th particle read
\begin{subequations}
\label{eq:distr}
\begin{align}
\vec{F}_{d,p} &=  \sum_{{\vec{r}_d} \in \mathcal{B}_p} \left[\rho^k(\vec{r}_d)+\rho^{\bar{k}}(\vec{r}_d) \right]\vec{u}(\vec{r}_d),\\
\vec{T}_{d,p} &= \sum_{{\vec{r}_d} \in \mathcal{B}_p} \left(\vec{r}_d-\vec{r}_p\right) \times \left\{ \left[\rho^k(\vec{r}_d)+\rho^{\bar{k}}(\vec{r}_d) \right]\vec{u}(\vec{r}_d) \right\},
\end{align}   
\end{subequations}
with the sum running over all the deleted fluid nodes, $\{\vec{r}_s\} \in \mathcal{B}_p$, of the $p$-th particle.

In the creation fluid node step, whenever the $p$-th particle leaves a lattice 
site located at $\vec{r}_c$ on its surface $\Sigma_p$, new fluid populations are initialized (see 
\cite{aidun1998direct,jansen2011bijels}) from the equilibrium distributions,
$f^{eq,k}_i(\bar{\rho}^k,\vec{u}_w)$, for the two $k$-th components with the velocity
of the particle wall, $\vec{u}_w$ estimated via Eq. \ref{UPI} and the fluid densities 
taken as the average values, $\bar{\rho}^k$ and $\bar{\rho}^{\bar{k}}$ respectively, of the
neighbouring nodes. As a consequence of the creation fluid step,
the associated linear and angular momenta, denoted $\vec{F}_{c,p}$ and $\vec{T}_{c,p}$, have to be subtracted from the $p$-th particle:
\begin{subequations}
\label{eq:creat}
\begin{align}
\vec{F}_{c,p} &=  \sum_{{\vec{r}_c} \in \Sigma_p} -\left[\rho^k(\vec{r}_c)+\rho^{\bar{k}}(\vec{r}_c) \right]\vec{u}_w(\vec{r}_c),\\
\vec{T}_{c,p} &= \sum_{{\vec{r}_c} \in \Sigma_p} \left(\vec{r}_c-\vec{r}_p\right) \times \left\{- \left[\rho^k(\vec{r}_c)+\rho^{\bar{k}}(\vec{r}_c) \right]\vec{u}_w(\vec{r}_c) \right\},
\end{align}   
\end{subequations}
with the sum running over all the created fluid nodes, $\{\vec{r}_c\} \in \Sigma_p$, of the $p$-th particle.

Inter-particle interactions are handled by an extension of
standard molecular dynamics algorithms.
In particular, LBsoft implements the same strategy adopted by Jensen and Harting \cite{jansen2011bijels} to model hard spheres contact by an Hertzian repulsive force term.
Thus, the repulsive contact force depends on the Hertzian potential according to the relation \cite{hertz1881contact}
\begin{equation}
\label{hertzpot}
\vec{F}_{Hertz,\,i}=
\begin{cases} 
      -\frac{\partial }{\partial |\vec{r}_{i\,j}|} \left[ K_{Hertz}\,\left(2 \, R - |\vec{r}_{i\,j}| \right)^{\frac{5}{2}} \right] \frac{\vec{r}_{i\,j}}{|\vec{r}_{i\,j}|}  & |\vec{r}_{i\,j}| < 2\,R \\
      0 & |\vec{r}_{i\,j}|\geq 2\,R 
\end{cases}
,
\end{equation}
where $K_{Hertz}$ is an elastic constant and $\vec{r}_{i\,j}=(\vec{r}_j-\vec{r}_i)$ the  distance between $i$-th and $j$-th particles of radius $R$. 
Further, LBsoft implements a lubrication correction to recover the fluid flow description between two particle surfaces, whenever their distance 
is below the lattice resolution. Following the indications given by Nguyen and Ladd \cite{nguyen2002lubrication}, the lubrication force reads
\begin{equation}
\label{libric}
\vec{F}_{lub,\,i}=
\begin{cases} 
      -6 \pi \nu \frac{R^4}{4R^2} \left(\frac{1}{h}-\frac{1}{h_n} \right)\left(\vec{\upsilon}_i -\vec{\upsilon}_j \right) \cdot \frac{\vec{r}_{i\,j}}{|\vec{r}_{i\,j}|}  & h < h_n \\
      \: \: 0 & h\geq h_n 
\end{cases}
,
\end{equation}
where $h=(|\vec{r}_{i\,j}|-2\,R)$ and $h_n$ is a cut off distance. Note that the term $(\vec{\upsilon}_i -\vec{\upsilon}_j)$ represents the relative velocity of the $i$-th particle with respect to the $j$-th one, so that a pair of perpendicularly colliding particles maximizes the repulsive lubrication force.

Finally, the total force and torque acting on the $p$-th particle reads
\begin{subequations}
\begin{align}
\vec{F}_p &= \vec{F}_{liq,\,p}+\vec{F}_{d,p}+\vec{F}_{c,p}+ \vec{F}_{Hertz,\,p}+\vec{F}_{lub,\,p} ,\\
\vec{T}_p &= \vec{T}_{liq,\,p}+\vec{T}_{d,p}+\vec{T}_{c,p} .
\end{align}   
\end{subequations}

We are now in a position to advance the particle position, speed $\vec{v}_p$ and angular momentum $\vec{\omega}_p$, according to Newton's equations of motion:

\begin{subequations}
\begin{align}
\frac{d \vec{r}_p}{d t} &=  \vec{v}_p,\\
m_p \frac{d \vec{v}_p}{d t} &=  \vec{F}_p,\\
I_p \frac{d \vec{\omega}_p}{d t} &= \vec{T}_p,
\end{align}   
\end{subequations}
where $m_p$ and $I_p$ are the particle mass and moment of inertia, respectively.

The rotational motion of particles requires the description of the local reference frame 
of the particle with respect to the space fixed frame. 
For instance, denoted the local reference frame by the symbol $\star$, 
an arbitrary vector $\vec{d}^{\star}$ is transformed from the local frame to the space fixed frame by the relation
\begin{equation}
\label{rotation}
\vec{d} = 
\underline{\underline{\mathbf{R}}} \: \vec{d}^{\star},
\end{equation}
where $\underline{\underline{\mathbf{R}}}$ denotes the rotational matrix to transform from the local reference frame to the space fixed frame.
In order to guarantee a higher computational efficiency, LBsoft exploits the quaternion, a four dimensional unit vectors, $\mathbf{q}=[q_0,q_1,q_2,q_3]^{T}$ in order to describe the orientation of the local reference frame of particles. In quaternion algebra\cite{altmann2005rotations}, the rotational matrix , $\underline{\underline{\mathbf{R}}}$, can be expressed as 
\begin{equation}
\underline{\underline{\mathbf{R}}}=\left(
\begin{array}{ccc}
                                q_0^2 + q_1^2 - q_2^2 - q_3^2 
                                        & 2(q_1 q_2 - q_0 q_3) 
                                        & 2(q_1 q_3 + q_0 q_2) \\
                                2(q_1 q_2 + q_0 q_3) 
                                        & q_0^2 - q_1^2 + q_2^2 - q_3^2 
                                        & 2 (q_2 q_3 - q_0 q_1) \\
                                2(q_1 q_3 - q_0 q_2) 
                                        & 2 (q_2 q_3 + q_0 q_1)
                                        & q_0^2 - q_1^2 - q_2^2 + q_3^2 \\
\end{array}
\right),
\end{equation}
whereas the transformation of a vector $\vec{d}^{\star}$ from the local frame to the space fixed frame
is obtained as $\vec{d} = 
\mathbf{q} \: \vec{d}^{\star}  \mathbf{q}^{-1}$,
and \textit{vice versa} as $\vec{d}^{\star} = 
\mathbf{q}^{-1}  \vec{d} \: \mathbf{q}$.

Exploiting the quaternion representation, the particles are advanced in time according to the
rotational leap-frog algorithm proposed by Svanberg \cite{svanberg1997research}:
\begin{subequations}
\label{eq:lf}
\begin{align}
\vec{v}_p \left( t+\frac{1}{2} \right) = \vec{v}_p \left( t-\frac{1}{2} \right)+ \frac {\vec{F}_p \left( t \right)}{m_p} , \label{eq:lf1} \\
\vec{r}_p \left(t + 1 \right) = \vec{r}_p \left( t \right) + \vec{v}_p \left( t+\frac{1}{2}  \right), \label{eq:lf2} \\
\vec{\omega}_p^{\star} \left(t+\frac{1}{2} \right) = \vec{\omega}_p^{\star} \left( t-\frac{1}{2} \right)+ \frac {\vec{T}_p^{\star} \left( t \right)}{I_p^{\star}} , \label{eq:lf3}\\
\mathbf{q}_p \left(t + 1 \right) = \mathbf{q}_p \left( t \right) + \dot{\mathbf{q}}_p \left( t+\frac{1}{2}  \right), \label{eq:lf4}
\end{align}   
\end{subequations}
where $\dot{\mathbf{q}}=[\dot{q}_0,\dot{q}_1,\dot{q}_2,\dot{q}_3]^{T}$ denotes the quaternion time derivative of the $p$-th  particle. Since the quaternion time derivative reads
\begin{equation}
\label{eq:quatder}
\left(
\begin{array}{cccc}
\dot{q}_0 \\
\dot{q}_1 \\
\dot{q}_2 \\
\dot{q}_3 \\
\end{array}
\right)= \frac{1}{2} \left(
\begin{array}{cccc}
q_0 
& -q_1
& -q_2
& -q_3 \\
q_1
& q_0
& -q_3
& q_2\\
q_2 
& q_3
& q_0 
& -q_1\\
q_3 
& -q_2
& q_1
& q_0\\
\end{array}
\right) \left(
\begin{array}{cccc}
0 \\
\omega^{\star}_x \\
\omega^{\star}_y \\
\omega^{\star}_z \\
\end{array}
\right),
\end{equation}
the reckoning of $\dot{\mathbf{q}}_p \left( t+\frac{1}{2}  \right)$ requires
the evaluation of $\mathbf{q}_p \left(t +\frac{1}{2} \right)$ which is not known at time
$t +\frac{1}{2}$. Svanberg \cite{svanberg1997research} overcomes
the issue exploiting Eq. \ref{eq:lf4} integrated at half time step, 
so obtaining the implicit equation 
\begin{equation}
\label{svanimpl}
\mathbf{q}_p \left(t + \frac{1}{2} \right) = \mathbf{q}_p \left( t \right) + \frac{1}{2} \dot{\mathbf{q}}_p \left( t+\frac{1}{2}  \right),
\end{equation}
which is solved by iteration, taking as a starting value 
$\mathbf{q}_p (t + \frac{1}{2}) = \mathbf{q}_p ( t ) + \frac{1}{2} \dot{\mathbf{q}}_p ( t )$ 
to estimate via Eq. \ref{eq:quatder} the time derivative $\dot{\mathbf{q}}_p( t+\frac{1}{2})$ . As the updated value of 
$\mathbf{q}_p (t + \frac{1}{2})$ is assessed by Eq. \ref{svanimpl}, 
the procedure is repeated until the term $\dot{\mathbf{q}}_p( t+\frac{1}{2})$ converges.

Note that the force, $\vec{F}_{liq,\,p}$, and torque, $\vec{T}_{liq,\,p}$
, exerted from the fluid on the particles 
are computed at half-integer times. Therefore, the total force and torque are assessed as
\begin{subequations}
\begin{align}
\begin{split}
\vec{F}_p(t) & = \frac{\vec{F}_{liq,\,p}(t-\frac{1}{2})+\vec{F}_{liq,\,p}(t+\frac{1}{2})}{2}+\vec{F}_{d,p}(t)+\vec{F}_{c,p}(t) +\\
& +\vec{F}_{Hertz,\,p}(t)+\vec{F}_{lub,\,p}(t), 
\end{split}  \\
\vec{T}_p(t) & = \frac{\vec{T}_{liq,\,p}(t-\frac{1}{2}) + \vec{T}_{liq,\,p}(t+\frac{1}{2})}{2}+\vec{T}_{d,p}(t)+\vec{T}_{c,p}(t) .
\end{align}   
\end{subequations}

The set of Equations \ref{eq:lf} takes into account the full many-body 
hydrodynamic interactions, since the forces and torques are computed with 
the actual flow configuration, as dictated by the presence of all $N$ 
particles simultaneously.

\section{Implementation}
LBsoft is implemented in Fortran 95 using modules to minimize code cluttering. 
In particular, the variables having in common the description of certain features 
(e.g., fluids and particles) or methods (e.g., time integrators) are rounded 
in different modules. 
The code adopts the convention of the explicit type declaration with
PRIVATE and PUBLIC accessibility attributes in order to decrease
error-proneness in
programming. Further, the arguments passed in calling sequences of
subroutines have defined intent.

LBsoft is written to remain portable with 
ease (multi-architecture) and readable (to allow code contribution), 
as much as possible taking into account the complexity of the underlying model

LBsoft has been developed, from the very beginning, considering, as target, parallel computing platforms.
The communication among computing nodes exploits the de-facto standard Message Passing 
Interface (MPI).
The communication in LBsoft is implemented in the module \texttt{version\_mod} contained  
in the \texttt{parallel\_version\_mod.f90} file.
Note that the module \texttt{module version\_mod} is also provided in a serial version located 
in the \texttt{serial\_version\_mod.f90} file, which can be easily selected in the compiling 
phase (see below).

LBsoft is completely open source, available at the 
public repository on GitHub: \textbf{www.github.com/copmat/LBsoft}
under the 3-Clause BSD License (BSD-3-Clause).

The repository is structured in three directories: \textit{source}, \textit{tests}, and \textit{execute}.

All the source code files are
contained in the \textit{source} directory.
The code does not require external libraries, 
and it may be compiled on any UNIX-like platform. 
To that purpose, the \textit{source} directory contains a
UNIX \texttt{makefile} to compile and link the code into the executable 
binary file with different compilers. 
The makefile includes a list of targets for several common
compilers and architectures (e.g., Intel Skylake, and Intel Xeon Phi) are 
already defined in the \texttt{makefile} with proper flags for the activation
of specific instruction sets, such as the 512-bit Advanced Vector Extensions (AVX-512). 
The list can be used 
by the command \enquote{\texttt{make target}}, where \texttt{target} is one of the options 
reported in Tab \ref{Tab:targets}. 
On Windows systems we advice the user to compile LBsoft under the
command-line interface Cygwin \cite{racine2000cygwin} or the Windows
subsystem for Linux (WSL).

The \textit{tests}
directory contains a set of test cases for code validation and results replication.
The test cases can also help the user to create new input files.
Further, the folder contains also a simple tool for running all of them and
checking results: this simplifies working with the code as it quickly detects
when a set of changes break working features. Along with code versioning
(using git and public repository GitHub), 
this can also lead to automatic search of problematic code changes 
through the bisection git capability.

Finally, the binary executable file
can be run in the \textit{execute} directory.

\begin{table}
\begin{centering}
\begin{tabular}{ll}
\hline 
\textbf{target:}  & \textbf{meaning:}\tabularnewline
\hline 
\hline 
gfortran & GNU Fortran compiler in serial mode.\tabularnewline
gfortran-mpi & GNU Fortran compiler in parallel mode with Open
Mpi library.\tabularnewline
intel & Intel compiler in serial mode.\tabularnewline
intel-mpi & Intel compiler in parallel mode with Intel Mpi
library.\tabularnewline
intel-openmpi & Intel compiler in parallel mode with Open Mpi
library.\tabularnewline
intel-mpi-skylake & Intel compiler in parallel mode with Intel Mpi library \tabularnewline
 & and flags for Skylake processor features (AVX-512 activated).\tabularnewline
intel-mpi-knl & Intel compiler in parallel mode with Intel Mpi library \tabularnewline
 & and flags for Xeon Phi processor features (AVX-512 activated).\tabularnewline
help & return the list of possible target choices\tabularnewline
\hline 
\end{tabular}
\par\end{centering}

\protect\caption{List of targets for several common parallel environments,
which can be used by the command "\texttt{make target}".}

\label{Tab:targets}
\end{table}

The source code is structured as follows: the \texttt{main.f90} file
implements the execution flow of LBsoft, gathering in sequence the
main stages, which are initialization, simulation, and
finalization. The initialisation stage includes the reading of the
input file \texttt{input.dat}, describing the simulation setup and the
physical parameters of the simulation. Further details on the input
parameters and directives are reported below in Section
\ref{sec:Description-of-files}. Moreover, the initialization stage
involves the allocation of main data arrays, as well as their
initialization. At this stage, the global communication among
computing nodes is set up, and the simulation mesh is decomposed in
sub-domains, each one assigned to a node.

The simulation stage solves both the lattice-Boltzmann equation and, if
requested, the particles evolution for the simulation time specified in the input file.
The \texttt{fluids\_mod.f90} and \texttt{particles\_mod.f90} files collect all
variable declarations, relative definitions, and procedures dealing 
with fluids and particles, respectively.
In the simulation stage, several statistical observables of the system are
computed at prescribed time steps by suitable routines of the 
\texttt{statistic\_mod.f90} file. 
The observables can be printed on the output file, \texttt{statdat.dat}, in several forms
defined by specific directives in the input file.
Further, the savings of fluid density and flow fields alongside with particle positions and
orientations can be activated in the input file and stored in different output formats.

Finally, the finalisation stage closes the global communication environment, 
deallocates previously allocated memory, and write a the restart files before 
the program closes successfully.

In the following Subsection, a few key points of the model implementation will be
outlined in more detail.

\subsection{Algorithmic details}
The time-marching implementation of the lattice-Boltzmann equation (Eq. \ref{eq:bgk})
can be realized by different approaches: the two-lattice, two-step, and fused swap
algorithms, to name a few \cite{pohl2003optimization,mattila2008comparison,wittmann2013comparison,succi2019towards}.
All these implementations manage the collision (Eq. \ref{eq:bgk1}) and streaming
(Eq. \ref{eq:bgk2}) steps following a different treatment of data dependence.
For instance, since the streaming step is not a local operation, it can be
implemented with a temporary copy in memory of old values for not
overwriting locations in the direction of $\vec{c}_i$,
or with a careful memory traversal in direction opposite to $\vec{c}_i$, thus avoiding
memory copies.

The two steps can be fused in one, for maximal reuse of data after fetching from 
memory and to avoid latencies in the streaming step (fused algorithm), 
at the cost of code complexity and increased level of difficulty 
in experimenting with different force types.
The interaction of the particle solver makes this fused approach even more challenging, 
since force on particles are needed at intermediate time steps with respect to LB times.

LBsoft implements the more modular form provided in the two-step algorithm \cite{mattila2008comparison}.
Indeed, the shift algorithm preserves separate subroutines or functions 
to compute the hydrodynamic variables, the equilibrium distributions, the collision step and 
the streaming step.
The highly modular approach of the shift algorithm allows a simpler coupling strategy 
to the particle solver.

The main variables of the code are the 19 distributions, the recovered physical variables 
(\( \rho \),\( \vec{u}\)) and a 3D field of booleans indicating whether the lattice point 
is fluid or not (a particle or an obstacle). 
The LB distributions are usually stored in two different memory layouts which are
mutually conflicting.
Regardless of the memory order (row-major like in C or column-major as in FORTRAN), the 
\texttt{npop} populations of the 3-d lattice of size \texttt{[nx,ny,nz]} can be stored as an  
array of structures (AoS) or as a structure of arrays
(SoA) \cite{succi2019towards}.

The chosen memory layout in LBsoft is the SoA, which proved to be more performing than the
AoS layout. In particular, the streaming step in the SoA reduces to a unitary stride. 
Hence, the SoA layout is optimal for streaming with a significant decrease in 
memory bandwidth issues.

In order to solve the lattice-Boltzmann equation, the algorithm
proceeds by the following sequence of subroutine calls:
\begin{enumerate}
    \item Recover density and velocity from populations (using Eq. \ref{eq:rho} and Eq. \ref{eq:vel});
    \item Compute the force term if necessary (e.g., Eq. \ref{shanchen});
    \item Compute populations after collision (using Eqs \ref{eq:bgk1} and \ref{eq:LEQ});
    \item Apply halfway bounce-back rules (e.g., Eq. \ref{eq:bb});
    \item Stream populations (using Eq. \ref{eq:bgk2}).
\end{enumerate}

Note that for the coupling forces between fluids, two subroutines deal with
the evaluation of the pseudo potential fields and their gradients, respectively, in 
order to assess Eq. \ref{shanchen}. 
Hence, the coupling force is added into the LB scheme as in Eq. \ref{eq:lbsbgk} 
to perform the collision step.
In the \textit{tests}, a dedicated test is present in the  \textit{3D\_Spinoidal} 
folder to simulate the separation of two fluids in two complete regions.

In the case of particle simulation, the sequence of subroutine calls is:
\begin{enumerate}
    \item Recover density and velocity from populations (using Eq. \ref{eq:rho} and Eq. \ref{eq:vel});
    \item Destroy and create fluid nodes according to moving particles and compute relative force correction terms (using Eqs \ref{eq:distr} and \ref{eq:creat});
    \item Compute the forces among particles (using Eqs \ref{hertzpot} and \ref{libric});
    \item Compute the LB force term (using Eq. \ref{soliddens} and \ref{shanchen2});
    \item Compute populations for eack component after LB collision (using Eqs \ref{eq:bgk1} and \ref{eq:LEQ});
    \item Apply the halfway bounce-back rule at particle surface and the relative force terms on particles (using Eqs \ref{eq:LADDFLBE} and \ref{eq:LADDFORCE});
    \item Evolve position and angular velocity of particles (using Eq. \ref{eq:lf});
    \item Stream populations for each component (using Eq. \ref{eq:bgk2}).
\end{enumerate}

It is worth to stress that the net force and torque exerted 
from the fluid on a particle (step 6 in the previous scheme) is obtained by summing over all 
wall sites of the surface particle as reported in Eqs \ref{FORCEPARTEQ} and \ref{TORQUEPARTEQ}.
These summations add several numerical terms of opposite signs, giving results having a very small
value in magnitude (close to zero if the surrounding fluid is at rest).
The finite numerical representation in floating-point arithmetic leads to loss of significance
in the sum operator due to the round-off error. The issue is observed in LBsoft, as
the summation order is modified, providing different results of the arithmetic operation up 
to the eighth decimal place in double-precision floating-point format.
Whenever the accuracy of the particle trajectories is essential, it is possible to compile
LBsoft with the preprocessor directive \texttt{QUAD\_FORCEINT} in order to perform these summations
in quadruple-precision floating-point format, recovering the loss of significance.

The motion of the colloidal particles (step 7 in the previous scheme) is numerically 
solved by a {\em leapfrog} approach. 
In particular, the evolution of the angular moment is treated by the {\em leapfrog}-like scheme
proposed by Svanberg \cite{svanberg1997research}, also referred to as mid-step implicit 
algorithm (see Eqs \ref{eq:lf} and \ref{svanimpl}).
The mid-step scheme was adopted after the implementation of different time advancing 
algorithms and after the evaluation of their numerical precision by a test case.
In particular, the test examined a rotating particle in a still fluid, measuring 
the numerical accuracy at which the code simulates a complete revolution with no energy drift.
The tested schemes were full-step \cite{fincham1992leapfrog}, the
mid-step \cite{svanberg1997research} and the simple leapfrog versions. 
The mid-step implicit leapfrog-like algorithm was found to be more consistent
with respect to other rotational leapfrog schemes, confirming previous observations in literature \cite{rozmanov2010robust,svanberg1997research}.

The inter-particle force computation exploits the neighbour lists, which track the particles
close to each $p$-th particle within a distance cut-off \texttt{rcut}.
The list is constructed by a combination of the link-cell algorithm with the 
Verlet list scheme \cite{frenkel2001understanding,hockney1988computer,auerbach1987special}.
Briefly, this strategy avoids the evaluation of the neighbour list at each time step, limiting 
the list update only to the cases in which a particle moves beyond a tolerance displacement, \texttt{delr},
given as an input parameter. As the lists need to be updated, the link-cell algorithm 
allows for the location of interacting particles with a linearly scaling cost, $O(N)$, being $N$ 
the total number of particles.

The comprehensive algorithmic organization for colloidal simulations provides a clear structure 
for the possible integration of future features. 
Indeed, the extensibility of the steps was overall preferred, preserving, as much as possible, the computational performances.

\subsection{Parallelization}

LBsoft is parallelized according to the single program multiple data paradigm \cite{darema2001spmd} using the Message Passing Interface (MPI) protocol. 
 In particular, the parallelization of the hydrodynamic part exploits the domain decomposition (DD) \cite{gropp1992parallel} of the 
simulation box using a one-, two- or three-dimensional block distribution among the MPI tasks. The selection of 
the block distribution is done at run-time (without recompiling the source code), and it can be tuned 
to obtain the best performance given the global grid dimension and the number of computing cores.
In the case of a cubic box, the best performance corresponds to the three-dimensional DD, 
which minimizes the surface to be communicated.
The current version of LBsoft features a halo having a width of one lattice spacing, which is necessary
to compute the local coupling force between the two fluid components.
Actually, the halo width could be easily increased for future algorithms.

Two main parallelization strategies, replicated data (RD) \cite{smith1991molecular,smith1993molecular} and 
domain decomposition \cite{pinches1991large,rapaport1991multi}, 
have been considered for the particle evolution in order to balance and equally 
divide the computation of particle dynamics among available cores, minimizing the 
amount of transmitted data among nodes.
The worst case scenario is a lattice cube with 20\% of the volume occupied by particles of 
radius in a range from 5.5 to 10 lattice nodes. Even in this extreme condition, 
the number of particles is in the order of 10K-20K, to be compared with lattice size of $1024^{3}$ (1G nodes).
Given the relative limited number of particles,
the Replicated Data (RD) parallelization strategy has been adopted in LBsoft.

The RD strategy generally increases the amount of transmitted data in global communications, nonetheless
the method is relatively simple to code and reasonably efficient for our cases of interest.
Within this paradigm, a complete replication of the particle physical variables (position, 
velocity, angular velocity) among the MPI tasks is performed. 
In particular, $\vec{r}_p$, $\vec{v}_p$ and $\vec{\omega}_p$ are allocated and maintained 
in all MPI tasks, but each MPI task evolves Eq. \ref{eq:lf} only for particles 
whose centers of mass lie in its sub-domain defined by the fluid partition.
Since each task evolves in time the particles residing in its local fluid sub-domain, 
the fluid quantities around each $p$-th particle (apart for the standard 1-cell halo needed 
for the streaming part of the LB scheme) does not require to be transferred to 
the task in charge of the particle.
In this way, the interactions between particles and surrounding fluid are 
completely localized without the need of exchanging data among tasks.
Only in the case of a particle crossing two or more sub-domains,
the net quantities of force and torque acting from the fluid on the particle 
(Eqs \ref{FORCEPARTEQ} and \ref{TORQUEPARTEQ}) are 
assessed at the cost of some global MPI reduction operations while performing the time advancing.
After the time step integration, new values of $\vec{r}_p$, $\vec{v}_p$ and $\vec{\omega}_p$ 
are sent to all tasks.
It is worth to highlight that the distribution of particles among MPI tasks 
(LB managed) may show unbalance issues,
particularly in particle aggregation processes where the particle density  
locally increases in correspondence of clustered structures.
Nonetheless, the cases studied so far do not show pathological effects due to workload unbalance.

\section{Description of input and output files \label{sec:Description-of-files}}

LBsoft needs an input file to setup the simulation.
The input file is free-form and case-insensitive. The input file is named 
\texttt{input.dat}, and it is structured in rooms. Each room starts with
the directive \texttt{[room "name"]} bracket by square brackets,
and it finishes with the directive \texttt{[end room]}. The \texttt{[room system]},
\texttt{[room fluid]}, and \texttt{[room particle]} compose
the list of possible choices.
The main room is the  \texttt{[room system]},
which specifies the number of time steps to be integrated, 
the size of the simulation box, the boundary conditions, the type of adopted domain
decomposition declares (1D, 2D, or 3D), and the directives invoking the output files. 
The other two rooms, \texttt{[room fluid]} and \texttt{[room particle]}, provide 
specification of various parameters for the two actors, respectively.
A list of key directives for each room is reported in \ref{append:directive}
for the benefit of potential LBsoft users.
Just few of these directives are mandatory.
A missed definition of any mandatory directive will call an error banner on the 
standard output.
In each room, the key directives can be written without a specific order.
Every line is read as a command sentence (record). 
Whenever a record begins with the symbol \# (commented), the
line is not processed. Each record is parsed in words (directives and additional 
keywords and numbers) with space characters recognized as separator elements.
Finally, the \texttt{[end]} directive marks the end of the input data.
In the current software version all the quantities have
to be expressed in lattice units.

In the case of particles simulation, a second input file, named \texttt{input.xyz},
has to be specified. The \texttt{input.xyz} file contains the specification of 
position, orientation, and velocity of particles. The file is structured as follows:
the first record is an integer indicating the particle number, while the second
may be empty or report the directive \texttt{read list} followed by appropriate 
symbolic strings, whose meaning is reported in \ref{append:inputkey}.
Hence, a sequence of $N$ lines follows with $N$ the number of particles.
Each line reports a character string identifying the particle type, the 
position coordinates, $x$ $y$ $z$, and, if the directive \texttt{read list} was
specified at the second record, a sequence of floating point numbers specifying the values
of the parameters following the directive \texttt{read list}.

Specific directives cause the writing of output files in different formats
(see \ref{append:directive}). Mainly, LBsoft prints on the standard output.
The reported information regards the code initialization, time-dependent data of the
running simulation, warning and error banners in case of problems
during the program execution.
The data reported in the standard output can be selected using 
the directive \texttt{print list} followed by appropriate symbolic strings.
We report in \ref{append:outputkey} the list of possible symbolic strings.   
Hence, the selected data are printed at the time interval indicated by the
directive \texttt{print every}. The next Section delivers few examples
of \texttt{input.dat} files.

The directive \texttt{print vtk} triggers the writing of the fluid density and flow
fields in a format compatible with the visualization toolkit (VTK) \cite{schroeder2004visualization}.
Whenever particles are present, their position and orientation are also written
in a VTK compatible format. The VTK format files can be read by common visualization programs
(e.g., ParaView \cite{ayachit2015paraview}).
Further, the directive \texttt{print xyz} provides the writing of 
a file \texttt{traj.xyz} with the particle positions written in XYZ
format and readable by suitable visualization 
programs (e.g. VMD-Visual Molecular Dynamics \cite{humphrey1996vmd}) to generate animations.

Finally, a simple binary output of relevant data can be activated by the key directive
\texttt{print binary}. Each Eulerian field will be saved
as a unique file in a simple binary matrix (or vector for scalar fields) resorting to the parallel input/output
subroutines available in the MPI library.
The last strategy is the fastest and more compact writing mode 
currently implemented in LBsoft.

\section{Test cases and performance}
\label{perfomance}

LBsoft provides a set of test cases in a dedicated directory.
Several physical scenarios verify the code functionalities 
testing all the implemented algorithms. Among them, there are the
Shear flow and the Pouseille flow in all the possible Cartesian directions
in order to test boundary conditions, forcing scheme, and, in general, the LB implementation
(\texttt{2D\_Shear} and \texttt{2D\_Poiseuille} sub-folders).
The spinoidal decomposition of two fluids in 3D inquires the functionality of 
Shan-Chen force algorithm.
Several tests containing a single particle serve to check the 
translation (\texttt{3D\_Particle\_pbc} folder) and rotational 
part of the leapfrog integrator (\texttt{3D\_Shear\_Particle} and \texttt{3D\_Rotating\_Particle} sub-folders).

These examples can be used to understand the capabilities of the
code, to become familiar with input parameters 
(looking at the \texttt{input.dat} files), and to check that code changes do not compromise basic behaviour (the
so called ``regression testing'').

Complex simulations require a code
able to deliver good performance with robust scaling properties. 
In order to assess the performance of LBsoft,
three cases were selected. 
The corresponding input files are briefly described below.

We probe the efficiency of the
implemented parallel strategy of LBsoft by quantitative estimators.
In order to measure the performance of the LB solver, the Giga 
Lattice Updates Per Second (GLUPS) unit is used. In particular, 
the definition of GLUPS reads:
\begin{equation}
\label{eq:glups}
\text{GLUPS}=\frac{L_x L_y L_z}{10^9 t_{\text{s}}},
\end{equation}
where $L_x$, $L_y$, and $L_z$ are the domain sizes in the
$x-$, $y-$, and $z-$ axis, and $t_{\text{s}}$ is the run (wall-clock) time (in seconds)
per single time step iteration. 
We also report the estimator MLUPSCore defined as the Lattice Updates Per Second and Per computing core:
\begin{equation}
\label{eq:mlups}
\text{MLUPSCore}=\frac{L_x L_y L_z}{10^6 t_{\text{s}}\, n_{p}},
\end{equation}
where $n_{p}$ denotes the number of processor cores.

Further, the speedup ($S_{p}$) is defined as 

\begin{equation}
S_{p}=\frac{T_{s}}{T_{p}},
\end{equation}

where $T_{s}$ stands for the CPU wall-clock time of the code executed on a single core, taken as a reference point,
and $T_{p}$ is the CPU wall-clock time of the code in parallel mode executed by
a number of processors equal to $n_{p}$ .

Hence, the parallel efficiency ($E_{p}$) is defined as

\begin{equation}
E_{p}=\frac{S_{p}}{n_{p}}.
\end{equation}

The benchmarks were carried out on two different computing platforms:
\begin{itemize}
    \item a cluster of servers each with 2 Intel Xeon Gold 6148 clocked at 2.8 GHz (20-core cpu), for a total of 40 cores per node, with 192 GB of DDR4, in the following labelled $(a)$;
    \item a cluster of servers each with 1 Intel Xeon Phi 7250 clocked at 1.4 GHz (68-core cpu), for a total of 68 cores per node, with 16 GB of MCDRAM and 96 GB of DDR4, in the following labelled $(b)$.
\end{itemize}
In all the following benchmarks, the LBsoft source code was compiled
using the Intel Fortran Compiler, version 2018, with the proper flags to activate the Advanced Vector Extensions 512-bit (AVX-512) instructions. 
The MPI library implementation provided by Intel was exploited to manage 
the parallel communications.
Hence, results and comments are below reported.

\subsection{Benchmarks\label{sub:singletest}}

In order to evaluate the computational performance 
of the lattice Boltzmann solver alone, we consider two cubic boxes of side 512 and 1024 lattice nodes
with periodic boundary along the three Cartesian axes. The boxes
are randomly filled with fluid mass density equal to 1.0 and zero velocity flow field 
(see benchmark test 1 in \ref{Tab:input-file-1}).

Table \ref{tab:case1} reports the wall-clock time measured on the
computing platform $(a)$ as a function of the cores number (strong
scaling) for both the box sizes alongside with the corresponding speed
up, $S_{p}$, and parallel efficiency, $E_{p}$.  We also reports the LB
performance metrics in terms of GLUPS, as defined in
Eq. \ref{eq:glups}.

\begin{table}[h!]
    \centering

    \begin{tabular}{|c|c|c|c|c|c|}
        \hline 
        $n_p$ &  $t_{\text{s}}$ (s) & {\scriptsize GLUPS} & {\scriptsize MLUPSCore} & $S_{p}$ & $E_{p}$ \\
        \hline 
128 & 0.294 & 0.46 & 3.59 & 1.00 & 100\% \\
256 & 0.152 & 0.88 & 3.44 & 1.92 & 96\% \\
512 & 0.079 & 1.70 & 3.32 & 3.68 & 92\% \\
1024 & 0.046 & 2.92 & 2.85 & 6,32 & 79\% \\
2048 & 0.025 & 5.37 & 2.62 & 11.52 & 72\% \\
        \hline 
        \hline
128 & 2.173 & 0.49 & 3.83 & 1.00 & 100\% \\
256 & 1.159 & 0.93 & 3.63 & 1.86 & 93\% \\
512 & 0.595 & 1.80 & 3.52 & 3.64 & 91\% \\
1024 & 0.321 & 3.34 & 3.26 & 6.72 & 84\% \\
2048 & 0.165 & 6.51 & 3.18 & 13.12 & 82\% \\
        \hline
    \end{tabular}

    \caption{Run (wall-clock) time in seconds
per single time step iteration, $t_{\text{s}}$, with corresponding GLUPS, MLUPSCore, speed up, $S_p$, and parallel efficiency, $E_p$, versus the number of computing cores, $n_p$ for a single component fluid simulation. From the top to bottom, results are reported for a cubic box of side $512$ and $1024$. Note that the parallel efficiency is reported in percentage.}
\label{tab:case1} 
\end{table}

Here, the parallel efficiency is higher than 90\% up to
512 cores (see Tab. \ref{tab:case1}).
Nonetheless, it is apparent that the efficiency decreases with the increase
of the communication burden (due to the high number of cores). 
This is commonly due to the communication latency, which deteriorates the performance 
whenever the workload assigned to each processor is not sufficiently high to offset 
the communication overheads. Hence, larger box sizes 
guarantees a high parallel efficiency also with a high number of computing cores, as shown
in Tab. \ref{tab:case1} for the larger box size case under investigation ($E_{p}\simeq 80 \%$ even on 2048 coresze).

The comparison between the two cases provides insights also on the
weak scaling properties of LBsoft.  In particular, it is observed an
increase in the wall-clock time of a factor close to eight scaling
from the smaller (512) up to the larger (1024) box size, confirming a
quasi linear weak scaling behaviour.
It is worth to highlight that the code performance are generally
comparable with that observed in literature
\cite{succi2019towards,schmieschek2017lb3d} following a similar
implementation strategy of keeping separated the collision and
streaming steps.

The entire LBsoft machinery is tested on 
the two component LB fluid model (see Subsection \ref{twocomp}) combined with the particle solver (Subsection \ref{refmd}) 
describing the colloids in a rapid demixing emulsion.
In order to assess the performance of the particle solver, we defined
three simulation setups with different numbers of particles.
In particular, three cubic boxes having a side of 245, 512 and 1024 lattice nodes were respectively 
filled with two fluids randomly distributed with initial fluid densities extracted from a 
Gaussian distribution of mean and standard deviation equal to 1.0 lu and 0.004 lu, respectively.
Hence, three values of the volume fraction occupied by particles are considered:
1\%, 10\% and 20\%, labeled $case \; 1$, $case \; 2$, and $case \; 3$, respectively.

The benchmark results will be also compared to the corresponding box sizes without particles, 
labeled $case \; 0$ (input file 2 reported in \ref{Tab:input-file-2}).
In all the cases, the Shan-Chen force term (see Subsection \ref{twocomp}) is 
activated by the key directive \texttt{force shanchen pair}, 
which serves also to set the coupling constant value equal to 0.65 lu.
The particles are assumed having all the same radius equal to 5.5 lu, and
are randomly distributed in the box. The initial particle velocity
was set equalt to zero by the key directive \texttt{initial temperature 0.d0} (see input-file-3 
of \ref{Tab:input-file-3}).
The particle wettability is set with an angle equal to $108^{\circ}$ 
with respect to the main axis $\vec{x}^{\star}$ in the local reference frame 
of each particle (key directive \texttt{force shanchen angle}). Note that the wettability is tuned by a 
smoothed switching function of the wettability angle, varying from -1 to 1
in a given range (in this case $10^{\circ}$) around the target value of $108^{\circ}$.
All the inter-particle force terms were computed
by means of neighbor's lists built with the shell distance cutoff equal to 12 lu, 
augmented by a tolerance value equal to 1 lu, set by the key directives 
\texttt{rcut} and \texttt{delr}, respectively.
The lubrication force is activated from a 2/3 lattice mutual distance
between two particles, as reported in previous simulations \cite{jansen2011bijels}. 
Further, the Hertzian repulsive potential starts as soon as two particles 
are closer than one lattice unit in order to avoid particle overlay events.
The particle mass was estimated as the weight corresponding to a particle 
made of silica \cite{herzig2007bicontinuous}. 

\begin{figure}[h]
\centerline{\resizebox{.6\linewidth}{!}{\includegraphics{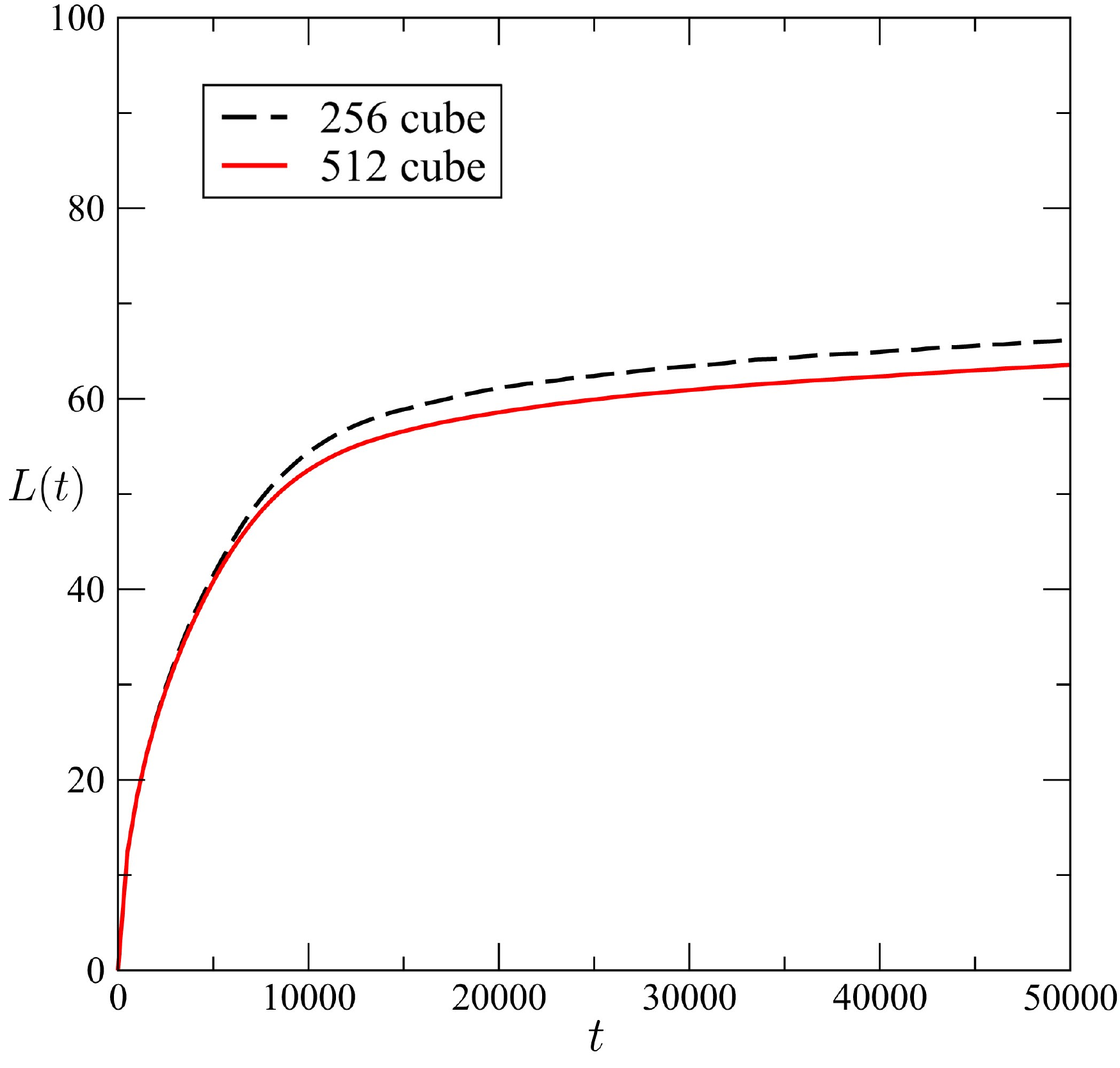}}}
\caption{Average domain size over time for the two systems of cubic side 256 and 512
at 20\% in particle volume fraction. After a rapid growth of the domain size to over 
60 lattice units during the first 20000 time steps, the $L(t)$ trend is stable 
close to the asymptotic value of 70 lattice units.}
\label{fig:trend_part}
\end{figure}

\begin{figure}[h]
\centerline{\resizebox{\linewidth}{!}{\includegraphics{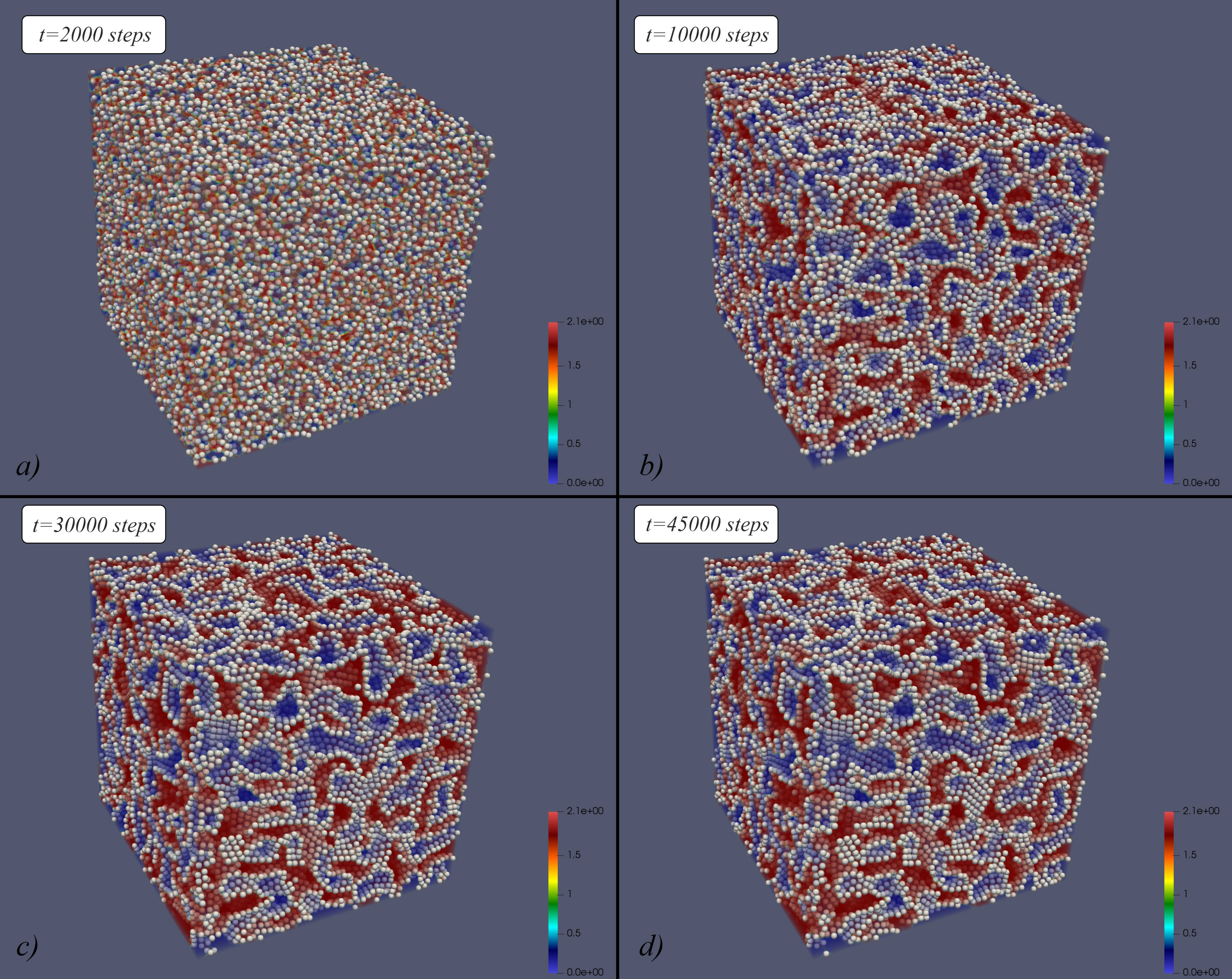}}}
\caption{3D visualization of the system with cubic box of side 512 lattice units described
in Figure \ref{fig:trend_part} at four different times. The particles are located close to
the interface between the fluid components. While the change in the particle configuration 
is evident from time steps 2000  up to 10000 (labelled $a$ and $b$, respectively), the domain 
growth is almost at rest in the following snapshots (snapshot $c$ and $d$) The density colour map is referred 
to the red fluid component.}
\label{fig:trend_snap}
\end{figure}

The test case describes the phase separation dynamics of two
immiscible liquids which is arrested in a rigid state by a jammed layer of colloidal 
particles located at the interface. The final metastable state is
a bi-continuous interfacially jammed emulsion gel, also referred to as a bi-jel \cite{cates2008bijels}.
The resulting dynamics can be described in terms of the average domain size measured along the time 
evolution, denoted $L(t)$ in the following text. Here, $L(t)$ is expected to show an asymptotic 
behaviour as the jammed layer arrests the domain growth, whereas 
in absence of colloidal particles $L(t)$ increases, showing a 
stretched exponential trend in the average domain size \cite{harting2007structural}. 
Several definitions of the average domain size were proposed in 
literature \cite{tiribocchi2011pattern,harting2007structural,kendon2001inertial,laradji1996molecular}. 
Here, the average size of domains is defined as the inverse of the first moment of 
the structure factor, $S(k,t)$, averaged on a sphere of radius $k$, being $k=|\vec{k}|$ the
modulus of the wave vector in Fourier space. Thus, $L(t)$ reads:
\begin{equation}
\label{Cluster_L_eq}
L(t) = 2 \pi  \frac{\displaystyle \int S(k,t)\mathrm{d}k } {\displaystyle \int k S(k,t) \mathrm{d}k}\: \text{,}
\end{equation}
being $S(k,t)\equiv  \langle \phi^{'}(\vec{k},t) \phi^{'}(-\vec{k},t)  \rangle$
with $\phi^{'}(\vec{k},t)$ the spatial Fourier transform of the fluctuations of the
order parameter $\phi^{'}=(\phi - \langle\phi\rangle)$. 
The order parameter $\phi$ is the phase field of the two components, 
called red and blue, $\phi = (\rho_r-\rho_b)/(\rho_r+\rho_b)$.

Figure \ref{fig:trend_part} shows the average domain size versus time, $L(t)$, for
the simulation box sizes of $256^3$ and $512^3$ lattice nodes with 5000 and 40000 particles, respectively,
corresponding to case (3) with 20\% in particle volume fraction.
Here, the $L(t)$ trend clearly shows an initial increase due to
a brief period of phase decomposition up to about 20000 time steps. 
After this initial jump, the asymptotic behaviour is observed for
both the box sizes with the converging value close to $\sim 70$
lattice units. The result is in agreement with similar observations in literature \cite{jansen2011bijels,stratford2005colloidal}, validating
the colloidal particle implementation in LBsoft.
The arrest of the phase demixing process is also evident by visualizing
the set of snapshots taken at different times (reported in Fig. \ref{fig:trend_snap}) for
the cubic box of side 512. In particular, Figure \ref{fig:trend_snap} clearly shows as the 
particles located at the fluid-fluid interface entrap the
demixing process into a metastable state, the bi-continuous jammed gel state.

\begin{table}[!h]
\label{tab:case3}
    \centering

    \begin{tabular}{|c|c|c|c|c|c|}
        \hline 
        $n_p$ &  $t_{\text{s}}$ (s) & {\scriptsize GLUPS} & {\scriptsize MLUPSCore} & $S_{p}$ & $E_{p}$ \\
        \hline
128 & 6.012 & 0.18 & 1.41 & 1.00 & 100\% \\
256 & 2.9 & 0.37 & 1.45 & 2.06 & 103\% \\
512 & 1.484 & 0.72 & 1.41 & 4.04 & 101\% \\
1024 & 0.784 & 1.37 & 1.34 & 7.60 & 95\% \\
        \hline
        \hline
128 & 6.4 & 0.17 & 1.33 & 1.00 & 100\% \\
256 & 3.63 & 0.30 & 1.17 & 1.76 & 88\% \\
512 & 2.12 & 0.51 & 1.00 & 3.04 & 76\% \\
1024 & 1.40 & 0.76 & 0.45 & 4.56 & 57\% \\
        \hline
        \hline 
128 & 8.42 & 0.13 & 1.02 & 1.00 & 100\% \\
256 & 4.79 & 0.22 & 0.86 & 1.76 & 88\% \\
512 & 2.77 & 0.39 & 0.76 & 3.04 & 76\% \\
1024 & 1.96 & 0.55 & 0.54 & 4.32 & 54\% \\
        \hline
        \hline 
128 & 10.53 & 0.10 & 0.78 & 1.00 & 100\% \\
256 & 6.00 & 0.18 & 0.70 & 1.76 & 88\% \\
512 & 3.63 & 0.30 & 0.59 & 2.92 & 73\% \\
1024 & 2.61 & 0.41 & 0.40 & 4.00 & 50\% \\
        \hline
    \end{tabular}

    \caption{Run (wall-clock) time, in seconds,
per single time step iteration, $t_{\text{s}}$, with corresponding GLUPS, MLUPSCore, speed up, $S_p$, and parallel efficiency, $E_p$, versus the number of computing cores, $n_p$ for a two component fluid in a cubic box of side $1024$ without
particles (case 0), and with three numbers of colloidal particles, $N$ (cases 1,2 and 3). From the top to the bottom, the box $1024^3$ was filled with $N=0$, $N=15407$, $N=154072$, and $N=308144$ (corresponding to a particle volume fraction equal to 0\%, 1\%, 10\% and 20\%, respectively). Note that the parallel efficiency is reported in percentage.}
\label{tab:test3}
\end{table}

\begin{figure}[!h]
\centerline{\resizebox{.7\linewidth}{!}{\includegraphics{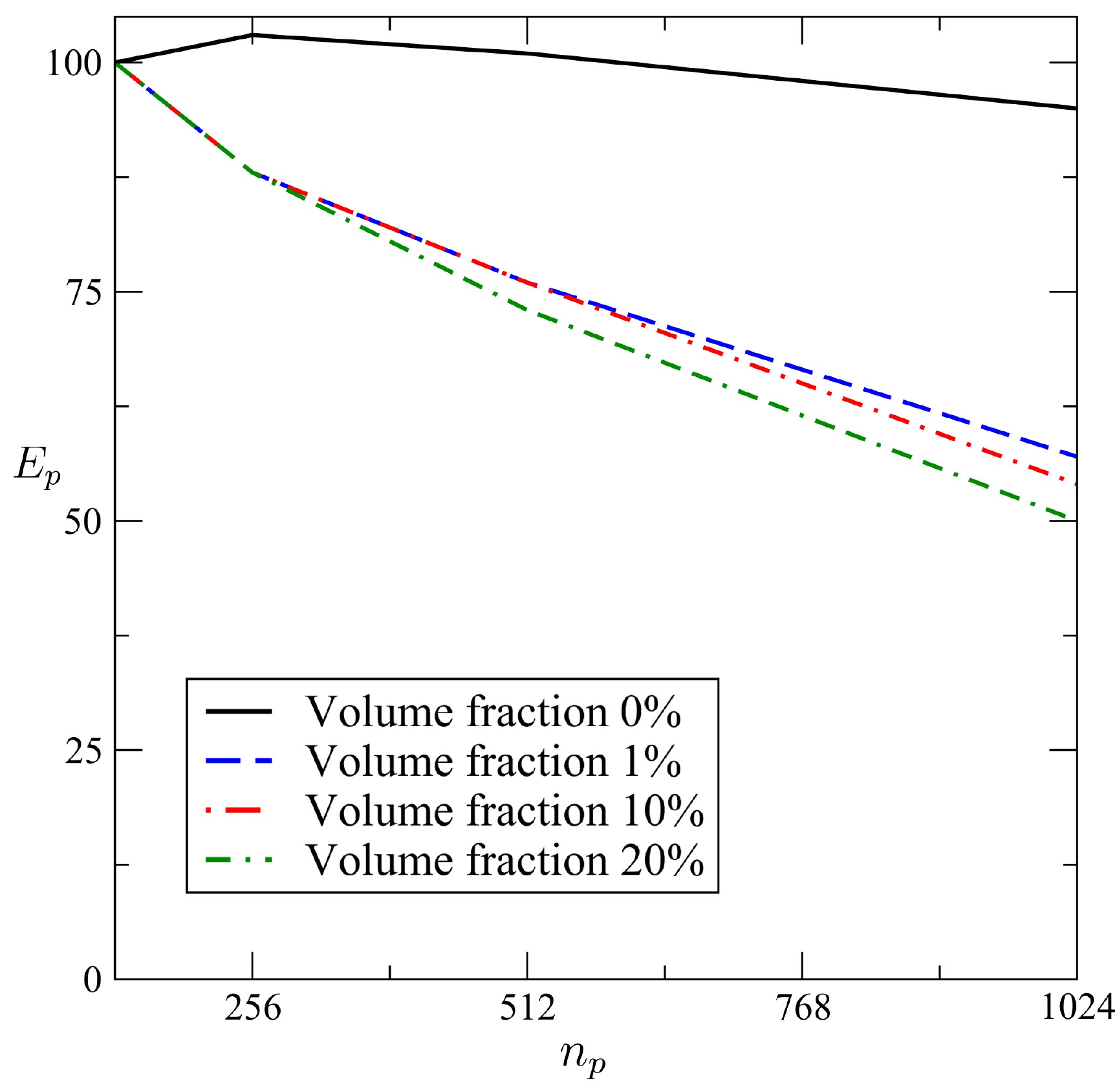}}}
\caption{Parallel efficiency, $E_p$, measured in a cubic box of $1024$ with different values in particle volume fraction versus the number of computing cores, $n_p$. Note that the $E_p$ is reported in percentage.}
\label{fig:paref}
\end{figure}

Table \ref{tab:case3} reports the strong scaling of wall-clock time per iteration step,
$t_{\text{s}}$, as a function of the core number, $n_p$, for the cubic box of side
1024 filled with different numbers of particles. 
The respective GLUPS and parallel efficiency, $E_{p}$, are also 
reported in Tab. \ref{tab:case3}. 
The measurements are carried out for all the three 
values of volume fraction under investigation (case 1, 2, and 3). 
In particular, the comparison among 
the three cases provides valuable information on the performance scaling as a function of the 
particle number in order to probe the weight of the MD part on the simulation time.

As a first observation, the measured wall-clock timings are about
2 and 3 times higher than the corresponding values measured in the
single-fluid case under the same conditions of cores number and box
size. The last finding is in agreement with previous observations
reported in the literature
\cite{succi2019towards,schmieschek2017lb3d}.

In Tab. \ref{tab:test3}, the simulations of $case \; 0$ without particles show 
wall-clock times which are about 2 and 3 higher than the corresponding values measured in the
single-fluid case under the same conditions of cores number and box
size. The last finding is in agreement with previous observations
reported in the literature
\cite{succi2019towards,schmieschek2017lb3d}.

Whenever the particle solver is activated, the parallel efficiency, $E_p$, remains above 50\% in all 
the three cases under investigation (see Fig. \ref{fig:paref}). As a matter of fact, although it is apparent 
that the particle solver degrades 
the performance of the entire simulation, that decrease in $E_p$ is
comparable to codes specifically designed for molecular dynamics simulations 
(e.g. LAMMPS \cite{plimpton1995fast}, DL\_poly \cite{todorov2006dl_poly_3}, etc.). 
For instance, a simulation of Lennard-Jones liquid of 30000 atoms shows in LAMMPS code 
a parallel efficiency, $E_p $, oscillating from 10\% up to 40\% on 512 cores depending 
on the computing architecture \cite{lammps}. 

Nonetheless, the analysis of the performance as a function of the number of particles 
shows a clear bottle-neck, going from 1\% to 20\% in the particle volume fraction. 
Here, the parallel efficiency is not scaling with the harder workload per core in 
the particle number, $N$. This is likely due to the Replicated Data strategy adopted 
in the implementation of the particle solver in LBsoft.
In order to recover the parallel efficiency in the particle scaled-size problem, 
further code tuning is ongoing. The most promising strategy involves 
a mixed OpenMP + MPI parallelization, in which the code utilises only a few 
MPI processes per node, and exploits all cores of a node with no explicit communication using OpenMP.  
This paradigm will reduce consistently the MPI global communications needed for 
the evolution of the particles with overall benefits in the parallel efficiency.

Finally, the code was also tested in the same conditions on a cluster of
servers based on Intel KNL processors, computing architecture (b),
courtesy of CINECA supercomputing center. Regardless of changes
introduced to let the Intel Fortran compiler provide a better
optimisation of the code, we observed a factor close to three of
performance degradation in all the conditions under investigations.
Apart the obvious lower clock frequency, the quite poor result may be motivated
by a sub-optimal exploitation of the limited MCDRAM size (16 GB) in the
Intel Xeon Phi processor, which could be improved using arbitrary
combinations of MPI and OpenMP threads within an hybrid parallel
implementation paradigm \cite{jarvis2017high}.

\subsection{New applications: confined bijels under shear\label{sub:confined}}

\begin{figure}[!h]
\centerline{\resizebox{0.7\linewidth}{!}{\includegraphics{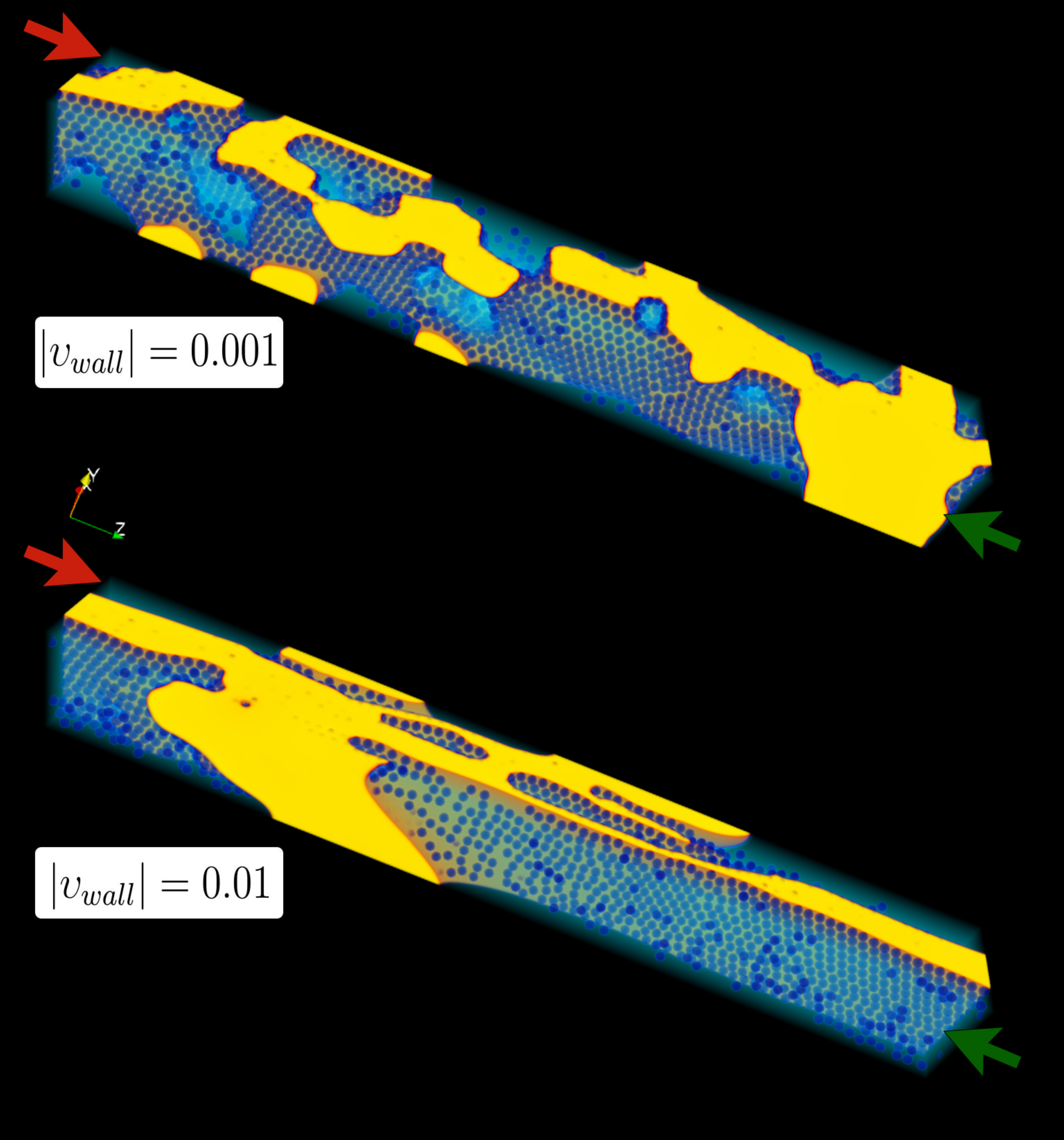}}}
\caption{From top to bottom, two snapshots of the system undergo a shear rate, $ |\upsilon_{wall}|$, of values $0.001$ and $0.01$, respectively. Both the images were collected after 400000 steps of time integration.}
\label{fig:confined}
\end{figure}

\begin{figure}[!h]
\centerline{\resizebox{0.7\linewidth}{!}{\includegraphics{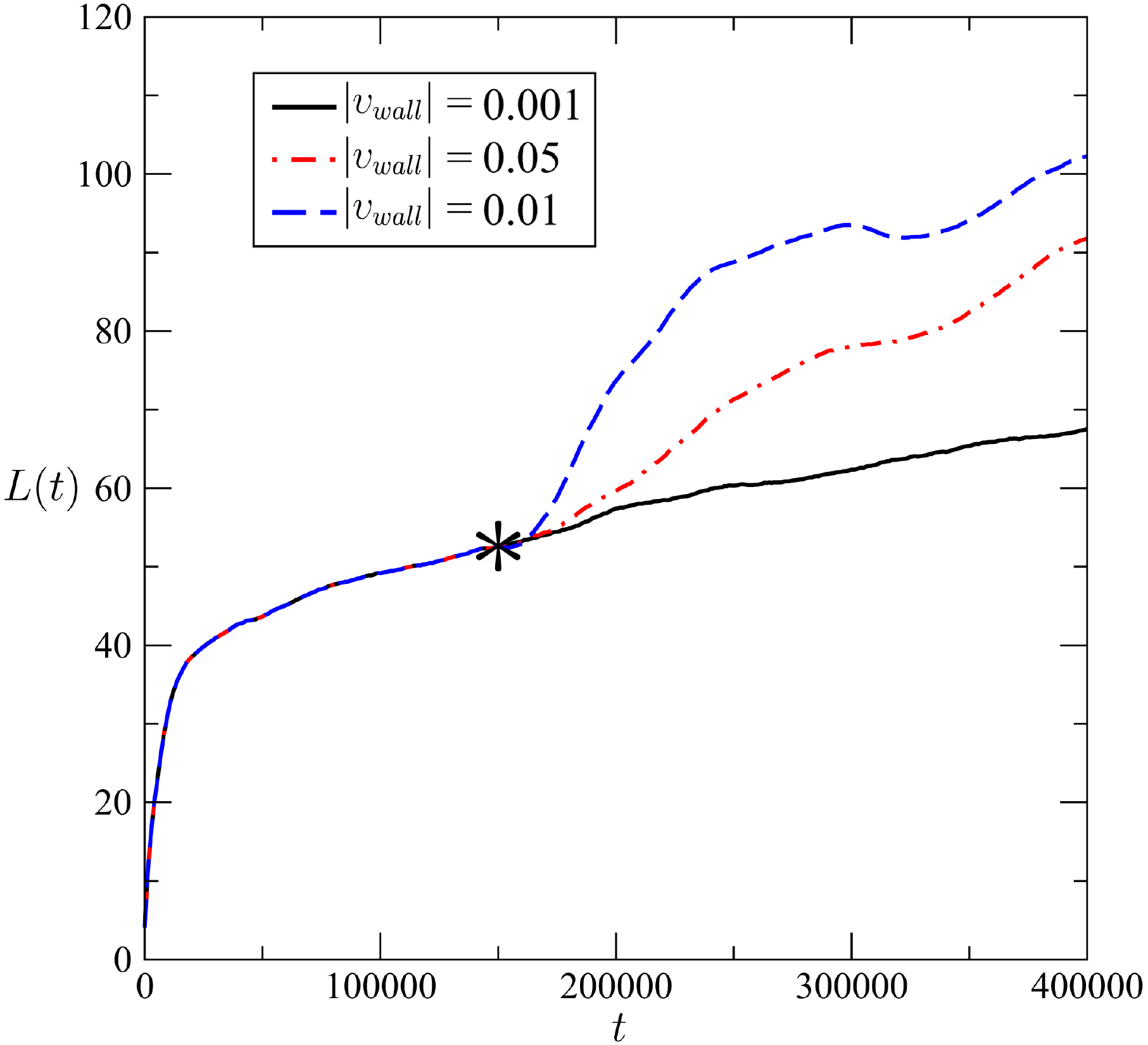}}}
\caption{Average domain size over time for an orthogonal box of sides 128, 128, and 1024
at 15\% in particle volume fraction. After a rapid growth of the domain size to over 
the initial 150000 time steps of equilibration, the system undergoes three different conditions of shear rate: $ |\upsilon_{wall}|=0.001$, $0.05$, and $0.01$.
The star symbol denotes the moment as the shear rate is applied on the bijel.}
\label{fig:confined_domain}
\end{figure}

LBsoft opens up the opportunity to investigate the rheology of bijels not only in terms
of bulk properties, but also under confinement. 
Leaving a detailed investigation to future publications, in the following we just wish to convey
the flavor of the possibilities offered by the code.
In particular, we report preliminary results of {\it confined} bijels under different conditions of shear rate. 

This framework is of utmost importance not only as a fundamental non-equilibrium process, but also
in view of numerous practical applications in material science, manufacturing, food processing, to name
but a few. As an example, a better control of the bijel formation process in microfluidic devices would greatly
benefit the rational design of the new bijel-based materials. 
Particularly notable is the case of 3D printing, where the possible application of the direct-write 
printing technique with bijels requires an in-depth knowledge of the effects of high shear rate 
as they occur in the micro-channels of the printing head.

The simulation setup discussed here involves an orthogonal box of sides $128 \times 128 \times 1024$ 
along the three Cartesian axes $x$, $y$, and $z$, respectively. 
The system is periodic along the $z$-axis (mainstream direction), while it is confined along the $x$ and $y$ directions. 
In particular, the two walls perpendicular to the $x$-axis move  at a constant velocity,$ |\upsilon_{wall}|$, oriented 
along the $z$-axis, with opposite directions (see Fig. \ref{fig:confined}). 
The walls perpendicular to the $y$-axis are set to no-slip boundary conditions. 

The box is filled by two fluids, randomly distributed with initial fluid densities extracted from a Gaussian distribution 
of mean and standard deviation $1.0$ and $0.004$ lattice units, respectively.
Colloidal spheres of radius $5.5$ are randomly distributed in the box, corresponding 
to a  particle volume fraction of  $0.15\%$.
The system is initially equilibrated for $150,000$ timesteps, in order to attain the bijel formation. 
Subsequently, three different shear rates, $ |\upsilon_{wall}| =0.001,0.005, 0.01$ respectively, are
applied to the system.  

In the absence of shear and confinement, it is well known that the effect of colloids is to segregate around
the interface, thereby slowing down and even arresting the coarse-graining of the binary mixture.
This has far-reaching consequences on the mechanical and rheological properties of the corresponding
materials. The application of a shear rate at the confining walls is expected to break up the fluid domains, 
thereby reviving the coarse-graining process against the blocking action of the colloids.
This is exactly what the simulations show, as reported Fig. \ref{fig:confined}, which
portrays two snapshots of the confined bijel configurations after $40,000$ timesteps.
  
In more quantitative terms, Fig. \ref{fig:confined_domain} reports the average domain 
size $L(t)$ for the three cases $ |\upsilon_{wall}|=0.001,0.005, 0.01$, as a function of time.
As one can appreciate, the smallest shear leaves the coarsening process basically
unaffected, while for $ |\upsilon_{wall}|=0.005, 0.01$, coarsening is clearly revived.

Many directions for future investigations can be envisaged: among others, i) a 
systematic study of the rheology of bijels under confinement, ii) the rheological
behaviour under cyclic shear applications (hysteresis),  iii) the inclusion of electrostatic 
interactions for the design of new materials based on polar bijels.

It hoped and expected that LBsoft may provide a valuable tool for the
efficient exploration  of the aforementioned topics and for the computational
designed of bijel-like based materials in general.

\section{Conclusion}

We have presented LBsoft, an open-source software specifically 
aimed at numerical simulations of colloidal systems. 
LBsoft is implemented in FORTRAN programming language and
permits to simulate large system sizes exploiting efficient parallel implementation.
In particular, the strong scaling behaviour of the code shows that
LBsoft delivers an excellent parallelization of the lattice-Boltzmann algorithm.
Further, whenever the particle integrator is combined with the fluid solver, LBsoft continues to show a
good performance in terms of scalability in both system size and number of processing cores.

In this work, the basic structure of LBsoft is reported along with the
main guidelines of the actual implementation. 
Furthermore, several examples have been reported 
in order to convey to the reader a flavor of the typical problems which 
can be dealt with using LBsoft. 
In particular, the simulations of bi-jel systems
demonstrate the capabilities of the present code to reproduce the complex dynamics of
the particles trapped at the jammed layers in a rapid de-mixing emulsion.

The LBsoft code is open source and completely accessible on the public GitHub repository. 
This is in line with the spirit of open source software, namely to promote the contribution 
of independent developers to the user community.

\section{Acknowledgments}
We would like to thank Prof. Anthony Ladd for illuminating discussions on rigid body models and precious suggestions on their implementations.
The software development process has received funding from the European
Research Council under the European Union's  Horizon 2020 programme /ERC Grant Agreement n.739964  ``COPMAT".
ML acknowledges the project ``3D-Phys" (PRIN 2017PHRM8X) from MIUR.

\appendix

\section{Directives of input file}
\label{append:directive}
\footnotesize
\begin{longtable}{ll}
\caption{Here, we report the list of directives available in LBsoft. Note
$i$, $f$, and $s$ denote an integer number, a floating point number,
and a string, respectively. Note that the input parameters should be expresses in lattice units.}
\tabularnewline
\endlastfoot
\hline 
\textbf{directive in system room:}  & \textbf{meaning:}\tabularnewline
\hline 
box $i_1$ $i_2$ $i_3$  &  dimension of the simulation box along the Cartesian axes\tabularnewline
bound cond $i_1$ $i_2$ $i_3$  & set boundary conditions along the Cartesian axes (0=wall,1=periodic)\tabularnewline
close time $f$ & Set job closure time to $f$ seconds\tabularnewline
diagnostic yes  & active the print of times measured in several subroutines of LBsoft code\tabularnewline
decompos type $i$  &  type of domain decomposition fo parallel jobs \tabularnewline
decompos dimen $i_1$ $i_2$ $i_3$  &  domain decomposition along the Cartesian axes\tabularnewline
job time $f$  & Set job time to $f$ seconds\tabularnewline
print list $s_{1}\,\ldots$  & print on terminal statistical data as indicated by symbolic strings\tabularnewline
print list every $i$  & print on terminal statistical data every $i$ steps\tabularnewline
print binary yes & activate the raw data print in binary format\tabularnewline
print binary every $i$  & print raw data in binary file format every $i$ steps (default: 100)\tabularnewline
print vtk yes & activate the output print in vtk files format\tabularnewline
print vtk every $i$  & print the output in vtk files format every $i$ steps\tabularnewline
print vtk list $s_{1}\,\ldots$  & print on vtk files the data indicated by symbolic strings\tabularnewline
print xyz yes & activate the particle print in XYZ file format\tabularnewline
print xyz every $i$  & print particle positions in XYZ file format every $i$ steps (default: 100)\tabularnewline
restart yes  & restart LBsoft from a previous job \tabularnewline
restart every $i$  & print the restart files every $i$ steps (default: 10000)\tabularnewline
seed $i$  & seed control to the internal random number generator\tabularnewline
steps $i$  &  number of time integration steps equal to $i$\tabularnewline
test yes  & activate an hash function for a deterministic initialization of LBsoft\tabularnewline
\hline 
\textbf{directive in LB room:}  & \textbf{meaning:}\tabularnewline
\hline 
bound open east type $i$  &  type of open boundary on east wall\tabularnewline
 & (1=Dirichlet,2=Neumann,3=Equilibrium)\tabularnewline
bound open west type $i$  &  type of open boundary on west wall\tabularnewline
bound open front type $i$  &  type of open boundary on front wall\tabularnewline
bound open rear type $i$  &  type of open boundary on rear wall\tabularnewline
bound open north type $i$  &  type of open boundary on north wall\tabularnewline
bound open south type $i$  &  type of open boundary on south wall\tabularnewline
bound open east dens $f_1$ $f_2$ &  density of open boundary on east wall\tabularnewline
bound open west dens $f_1$ $f_2$ &  density of open boundary on west wall\tabularnewline
bound open front dens $f_1$ $f_2$ &  density of open boundary on front wall\tabularnewline
bound open rear dens $f_1$ $f_2$ &  density of open boundary on rear wall\tabularnewline
bound open north dens $f_1$ $f_2$ &  density of open boundary on north wall\tabularnewline
bound open south dens $f_1$ $f_2$ &  density of open boundary on south wall\tabularnewline
bound open east veloc $f_1$ $f_2$ $f_3$ &  velocity  vector of open boundary on east wall\tabularnewline
bound open west veloc $f_1$ $f_2$ $f_3$ &  velocity  vector of open boundary on west wall\tabularnewline
bound open front veloc $f_1$ $f_2$ $f_3$ &  velocity  vector of open boundary on front wall\tabularnewline
bound open rear veloc $f_1$ $f_2$ $f_3$ &  velocity  vector of open boundary on rear wall\tabularnewline
bound open north veloc $f_1$ $f_2$ $f_3$ &  velocity  vector of open boundary on north wall\tabularnewline
bound open south veloc $f_1$ $f_2$ $f_3$ &  velocity  vector of open boundary on south wall\tabularnewline
component $i$  &  number of fluid components equal to $i$\tabularnewline
dens mean $f_1$ $f_2$ &  mean values of the initial fluid densities\tabularnewline
dens stdev $f_1$ $f_2$ &  standard deviation values of the initial fluid densities\tabularnewline
dens back $f_1$ $f_2$ &  background values of the initial fluid densities\tabularnewline
dens rescal $i$ $f$ & rescale the mass density every $i$ steps if deviating of $f$\tabularnewline  
 & from the initial value taken as a reference point\tabularnewline
dens ortho $i$ $f_n : n\in[1,\dots,8]$ &  orthogonal object numbered $i$ with extremes $f_n : n\in[1,\dots,6]$ and\tabularnewline
 & densities $f_7$ and $f_8$ in the box\tabularnewline
dens sphere $i$ $f_n : n\in[1,\dots,7]$ &  spherical object numbered $i$ with center $f_n : n\in[1,\dots,3]$,\tabularnewline
 & radius and width $f_4$ and $f_5$ and densities $f_6$ and $f_7$\tabularnewline
dens gauss & set a gaussian distribution of the initial fluid densities\tabularnewline
dens uniform & set an uniform distribution of the initial fluid densities\tabularnewline
dens special & set a special definition of the initial fluid densities by\tabularnewline
 & a set of objects defined in input file\tabularnewline
force ext $f_1$ $f_2$ $f_3$ & external force along the Cartesian axes\tabularnewline
force shanc pair $f$ &  coupling constant of Shan-Chen force between the fluids\tabularnewline
force cap $f$ & set a capping value for the total force terms\tabularnewline
veloc mean $f_1$ $f_2$ $f_3$ &  mean initial fluid velocities along the Cartesian axes\tabularnewline
visc $f_1$ $f_2$ & viscosity of the fluids\tabularnewline
tau $f_1$ $f_2$ & relaxation time of the fluids for the BGK collisional\tabularnewline
wetting mode $i$ $f_1$ $f_2$ & model of wetting wall and respective constants\tabularnewline
 & (0=averaged density,1=constant density)\tabularnewline
\hline 
\textbf{directive in MD room:}  & \textbf{meaning:}\tabularnewline
\hline 
delr $f$ & set Verlet neighbour list shell width to $f$\tabularnewline
densvar $f$ & density variation for particle arrays\tabularnewline
field pair wca $i_1$ $i_2$ $f_1$ $f_2$ & pair of particle types, force constant and\tabularnewline
 & minimum distance of the Weeks-Chandler-Andersen potential\tabularnewline
field pair lj $i_1$ $i_2$ $f_1$ $f_2$ & pair of particle types, force constant and\tabularnewline
 & minimum distance of the Lennard-Jones potential\tabularnewline
field pair hz $i_1$ $i_2$ $f_1$ $f_2$ & pair of particle types, force constant,\tabularnewline
 & minimum distance and capping distance of the Hertzian potential\tabularnewline
force ext $f_1$ $f_2$ $f_3$ & external particle force along the Cartesian axes\tabularnewline
force shanc angle $f_1$ $f_2$ & inflection and width of the switching function\tabularnewline
 & tuning the particle wettability\tabularnewline
force shanc part $f_1$ $f_2$ & constants of wetting wall on particles\tabularnewline
force cap $f$ & set a capping value for the total particle force terms\tabularnewline
lubric yes $f_1$ $f_2$ $f_3$ & activate lubrication force with rescaling constant,\tabularnewline
 &  minimum distance and capping distance\tabularnewline
init temperat $f$ & amplitude of Maxwell distribution of initial particle velocity\tabularnewline
mass $i$ $f$ & set mass $f$ to the particle of type $i$\tabularnewline
moment fix $i$ & remove the linear momentum every $i$ steps\tabularnewline
particle yes & activate the particle part of LBsoft\tabularnewline
particle type $i$ $s$ $\dots$& number of particle types labelled by symbolic strings\tabularnewline
rcut $f$ & set required forces cutoff to $f$\tabularnewline
rotate yes & activate the rotation of the particles\tabularnewline
rotate yes & activate the rotation of the particles\tabularnewline
side wall const $f$ & constant of soft harmonic wall\tabularnewline
side wall dist $f$ & location of the minimum of soft harmonic wall\tabularnewline
torque ext $f_1$ $f_2$ $f_3$ & external particle torque along the Cartesian axes\tabularnewline
\hline 
\label{tab:inputlist}  & \tabularnewline
\end{longtable}
\normalsize

\section{List of keys of input particle file}
\label{append:inputkey}

\begin{longtable}{ll}
\caption{List of symbolic string keys is reported with their corresponding meanings
for the particles input file named \texttt{input.xyz}.}
\tabularnewline
\endlastfoot
\hline 
\textbf{keys:}  & \textbf{meaning:}\tabularnewline
\hline 
\hline 
mass   & mass of the particle\tabularnewline
vx     & particle velocity along x\tabularnewline
vy     & particle velocity along y\tabularnewline
vz     & particle velocity along z\tabularnewline
ox     & particle angular velocity along x\tabularnewline
oy     & particle angular velocity along y\tabularnewline
oz     & particle angular velocity along z\tabularnewline
rad    & radius of the spherical particle\tabularnewline
radx   & radius of the particle along x\tabularnewline
rady   & radius of the particle along y\tabularnewline
radz   & radius of the particle along z\tabularnewline
\hline 
\label{tab:inputlist}  & \tabularnewline
\end{longtable}

\section{List of keys of output observables}
\label{append:outputkey}

\begin{longtable}{ll}
\caption{In LBsoft a set of instantaneous and statistical data are available
to be printed by selecting the corresponding key. Here, the list of
symbolic string keys is reported with their corresponding meanings.
All the quantities must be expressed in lattice units, if not 
explicitly declared in Table.}
\tabularnewline
\endlastfoot
\hline 
\textbf{keys:}  & \textbf{meaning:}\tabularnewline
\hline 
\hline 
t     & time\tabularnewline
dens1 & mean fluid density of first component\tabularnewline
dens2 & mean fluid density of second component\tabularnewline
maxd1 & max fluid density of first component\tabularnewline
maxd2 & max fluid density of second component\tabularnewline
mind1 & min fluid density of first component\tabularnewline
mind2 & min fluid density of second component\tabularnewline
maxvx & max fluid velocity of fluid mix along x\tabularnewline
minvx & min fluid velocity of fluid mix along x\tabularnewline
maxvy & max fluid velocity of fluid mix along y\tabularnewline
minvy & min fluid velocity of fluid mix along y\tabularnewline
maxvz & max fluid velocity of fluid mix along z\tabularnewline
minvz & min fluid velocity of fluid mix along z\tabularnewline
fvx   & mean fluid velocity along x\tabularnewline
fvy   & mean fluid velocity along y\tabularnewline
fvz   & mean fluid velocity along z\tabularnewline
engkf & kinetic energy of fluid\tabularnewline
engke & kinetic energy of particle\tabularnewline
engcf & configuration energy\tabularnewline
engrt & rotational energy\tabularnewline
engto & total energy\tabularnewline
intph & fluid interphase volume fraction\tabularnewline
rminp & min pair distance between particles\tabularnewline
tempp & particle temperature as ratio of KbT\tabularnewline
maxpv & max particle velocity\tabularnewline
pvm   & mean particle velocity module\tabularnewline
pvx   & mean particle velocity along x\tabularnewline
pvy   & mean particle velocity along y\tabularnewline
pvz   & mean particle velocity along z\tabularnewline
pfm   & mean particle force module \tabularnewline
pfx   & mean particle force along x\tabularnewline
pfy   & mean particle force along y\tabularnewline
pfz   & mean particle force along z\tabularnewline
cpu   & time for every print interval in seconds\tabularnewline
cpur  & remaining time to the end in seconds\tabularnewline
cpue  & elapsed time in seconds\tabularnewline
\hline 
\label{tab:outputlist}  & \tabularnewline
\end{longtable}

\section{Benchmark input files}
\label{append:test1}

\begin{table}[h!]
\begin{centering}
\begin{tabular}{l}
\hline 
\textbf{Benchmark test 1.} Input file of single fluid test.\tabularnewline
\hline 
\hline 
\texttt{[ROOM SYSTEM]}\tabularnewline
\texttt{box 512 512 512}\tabularnewline
\texttt{steps 5000}\tabularnewline
\texttt{boundary condition 1 1 1}\tabularnewline
\texttt{decomposition type 7}\tabularnewline
\texttt{print list maxd1 mind1 maxvx maxvy maxvz}\tabularnewline
\texttt{print every 10}\tabularnewline
\texttt{test yes}\tabularnewline
\texttt{[END ROOM]}\tabularnewline
\texttt{[ROOM LB]}\tabularnewline
\texttt{components 1}\tabularnewline
\texttt{density gaussian}\tabularnewline
\texttt{density mean 1.0}\tabularnewline
\texttt{density stdev 1.d-4}\tabularnewline
\texttt{velocity mean 0.d0}\tabularnewline
\texttt{fluid tau 1.d0}\tabularnewline
\texttt{[END ROOM]}\tabularnewline
\texttt{[END]}\tabularnewline
\hline 
\end{tabular}
\par\end{centering}

\label{Tab:input-file-1}
\end{table}

\begin{table}[h!]
\begin{centering}
\begin{tabular}{l}
\hline 
\textbf{Benchmark test 2.} Input file of two fluid test.\tabularnewline
\hline 
\hline 
\texttt{[ROOM SYSTEM]}\tabularnewline
\texttt{box 512 512 512}\tabularnewline
\texttt{steps 5000}\tabularnewline
\texttt{boundary condition 1 1 1}\tabularnewline
\texttt{decomposition type 7}\tabularnewline
\texttt{print list maxd1 mind1 maxvx maxvy maxvz}\tabularnewline
\texttt{print every 10}\tabularnewline
\texttt{test yes}\tabularnewline
\texttt{[END ROOM]}\tabularnewline
\texttt{[ROOM LB]}\tabularnewline
\texttt{components 2}\tabularnewline
\texttt{density gaussian}\tabularnewline
\texttt{density mean 1.0 1.0}\tabularnewline
\texttt{density stdev 1.d-4 1.d-4}\tabularnewline
\texttt{velocity mean 0.d0 0.d0}\tabularnewline
\texttt{fluid tau 1.d0 1.d0}\tabularnewline
\texttt{force shanchen pair 0.65d0}\tabularnewline
\texttt{[END ROOM]}\tabularnewline
\texttt{[END]}\tabularnewline
\hline 
\end{tabular}
\par\end{centering}

\label{Tab:input-file-2}
\end{table}

\begin{table}[h!]
\begin{centering}
\begin{tabular}{l}
\hline 
\textbf{Benchmark test 3.} Input file of bi-jel test.\tabularnewline
\hline 
\hline 
\texttt{[ROOM SYSTEM]}\tabularnewline
\texttt{box 512 512 512}\tabularnewline
\texttt{steps 5000}\tabularnewline
\texttt{boundary condition 1 1 1}\tabularnewline
\texttt{decomposition type 7}\tabularnewline
\texttt{print list maxd1 mind1 maxvx maxvy maxvz}\tabularnewline
\texttt{print every 10}\tabularnewline
\texttt{test yes}\tabularnewline
\texttt{[END ROOM]}\tabularnewline
\texttt{[ROOM LB]}\tabularnewline
\texttt{components 2}\tabularnewline
\texttt{density gaussian}\tabularnewline
\texttt{density mean 1.0 1.0}\tabularnewline
\texttt{density stdev 1.d-4 1.d-4}\tabularnewline
\texttt{velocity mean 0.d0 0.d0}\tabularnewline
\texttt{fluid tau 1.d0 1.d0}\tabularnewline
\texttt{force shanchen pair 0.65d0}\tabularnewline
\texttt{[END ROOM]}\tabularnewline
\texttt{[ROOM MD]}\tabularnewline
\texttt{particle yes}\tabularnewline
\texttt{particle type 1}\tabularnewline
\texttt{C}\tabularnewline
\texttt{rotate yes}\tabularnewline
\texttt{lubric yes 0.1d0 0.67d0 0.5d0}\tabularnewline
\texttt{densvar 5.d0}\tabularnewline
\texttt{rcut 12.d0}\tabularnewline
\texttt{delr 1.0d0}\tabularnewline
\texttt{shape spherical 1 5.5d0}\tabularnewline
\texttt{field pair hz 1 1 20.d0 12.d0 11.d0}\tabularnewline
\texttt{mass 1 472.d0}\tabularnewline
\texttt{initial temperature 0.0}\tabularnewline
\texttt{force shanchen angle 108.d0 10.d0}\tabularnewline
\texttt{force shanchen particel 0.1d0 0.1d0}\tabularnewline
\texttt{[END ROOM]}\tabularnewline
\texttt{[END]}\tabularnewline
\hline 
\end{tabular}
\par\end{centering}

\label{Tab:input-file-3}
\end{table}

\clearpage


\end{document}